\documentstyle[11pt]{article}

\font\fr=eufm10 scaled \magstep 1 %(caracteres goticos)
\font\es=msbm11                  %(caracteres ``doble barra´´)
\newtheorem{teor}{Theorem}
\newtheorem{prop}{Proposition}
\newtheorem{corol}{Corollary}

\newtheorem{definition}{Definition}
\def\beq{\begin{equation}}
\def\eeq{\end{equation}}
\def\bea{\begin{eqnarray}}
\def\eea{\end{eqnarray}}
\def\beann{\begin{eqnarray*}}
\def\eeann{\end{eqnarray*}}
\def\ben{\begin{enumerate}}
\def\een{\end{enumerate}}
\def\bit{\begin{itemize}}
\def\eit{\end{itemize}}
\def\dst{\(\displaystyle}
\def\derpar#1#2{\frac{\partial{#1}}{\partial{#2}}}

\def\feble#1{\mathrel{\mathop =\limits_{#1}}}
\def\feblesim#1{\mathrel{\mathop\sim\limits_{#1}}}

\def\moment#1#2#3{{#1}_{#2}, \ldots, {#1}_{#3}}

\def\qed{\ifvmode\removelastskip\fi
{\unskip\nobreak\hfil\penalty50\hbox{}\nobreak\hfil
\hbox{\vrule height1.2ex width1.2ex}\parfillskip=0pt
\finalhyphendemerits=0 \par\smallskip}}
\def\vf{\mbox{\fr X}}
\def\df{{\mit\Omega}}
\def\Lag{{\cal L}}

\def\d{{\rm d}}
\def\Real{\mbox{\es R}}

\def\inn{\mathop{i}\nolimits}
\def\Tan{{\rm T}}

\def\ls{(J^1E,\Omega_\Lag )}
\def\hs{(J^{1*}E,\Omega_h)}
\def\hso{(J^{1*}E,{\cal P},\Omega_h^0)}
\def\lag{\pounds}
\def\Cinfty{{\rm C}^\infty}
\def\proof{( {\sl Proof} )\quad}
\def\tabaddress#1{{\small\it\begin{tabular}[t]{c}#1 \\[1.2ex]\end{tabular}}}
\def\tabaddress#1{{\small\it\begin{tabular}[t]{c}#1 \\[1.2ex]\end{tabular}}}
\def\UPCMAT{Departamento de Matem\'atica Aplicada y Telem\'atica\\
   Campus Norte UPC, M\'odulo C-3\\
   C/ Jordi Girona 1-3,\\
   E-08034 Barcelona. Spain}

\title{ON THE CONSTRUCTION OF ${\cal K}$-OPERATORS IN FIELD THEORIES AS
       SECTIONS ALONG LEGENDRE MAPS}
\author{\sc Arturo Echeverr\'ia-Enr\'iquez,
 \\
   \tabaddress{\UPCMAT}
 \\
{\sc Jes\'us Mar\'\i n-Solano \thanks{{\bf e}-{\it mail}:
 JMARIN@ECO.UB.ES}},
 \\
 \tabaddress{Departamento de Matem\'atica Econ\'omica,
  Financiera y Actuarial, UB\\
  Av. Diagonal 690. E-08034 Barcelona. Spain}
   \\
{\sc Miguel C. Mu\~noz-Lecanda\thanks{{\bf e}-{\it mail}: MATMCML@MAT.UPC.ES}},
{\sc Narciso Rom\'an-Roy\thanks{{\bf e}-{\it mail}: MATNRR@MAT.UPC.ES}},
   \\
   \tabaddress{\UPCMAT}}
\date{Acta Applicandae Mathematicae {\bf 77}(1) (2003) 1-40}

\begin{document}
\maketitle
\thispagestyle{empty}

   \begin{abstract}
The ``time-evolution $K$-operator'' (or ``relative Hamiltonian vector field'')
in mechanics is a powerful tool 
which can be geometrically defined as a vector field along the
Legendre map. It has been extensively used by several authors
for studying the structure and properties of the dynamical systems
(mainly the non-regular ones), such as the relation between the
Lagrangian and Hamiltonian formalisms, constraints,
and higher-order mechanics.

This paper is devoted to defining a generalization of this operator
for field theories, in a covariant formulation.
In order to do this, we use sections along maps,
in particular multivector fields (skew-symmetric contravariant
tensor fields of order greater than 1), jet fields and connection forms
along the Legendre map.
As a relevant result, we use these geometrical objects
to obtain the solutions of the Lagrangian and Hamiltonian field equations,
and the equivalence among them (specially for non-regular field theories).
\end{abstract}


\bigskip
\bigskip
\vbox{\raggedleft AMS s.\,c.\,(2000):
 51P05, 53C05, 53C80, 55R10, 58A20, 58A30, 70S05.\\
PACS (1999): 02.40.Hw, 02.40.Vh, 11.10.Ef, 11.10 Kk}\null

\clearpage

\section{Introduction}

The so-called {\sl time-evolution $K$-operator} in mechanics
(also known by some authors as the {\sl relative Hamiltonian vector field})
is a tool which has mainly been developed
in order to study the Lagrangian and Hamiltonian formalisms
for singular mechanical systems and their equivalence.
It was first introduced in a non-intrinsic way in
\cite{BGPR-eblhf} as an ``evolution operator'' to connect both formalisms,
as a refinement of the technique used in \cite{Ka-82}.
This operator was later defined geometrically
 in two different but equivalent ways
\cite{CL-K}, \cite{GP-K} for autonomous dynamical systems,
and in \cite{CFM-95} for the non-autonomous case.
In \cite{GP-K}, a further different geometric construction
is given, using a canonical map introduced by Tulczyjew \cite{Tu-76}.

The following is a summary of relevant results obtained using this operator:
\bit
\item
The equivalence between the Lagrangian and Hamiltonian formalisms is proved
by means of this operator in the following way: there is a bijection between the
sets of solutions of Euler-Lagrange equations and Hamilton equations, even though
the dimensions of the final constraint submanifold in both formalisms are not
the same, in general (see \cite{BGPR-eblhf}, \cite{GP-ggf}).
\item
The complete classification of constraints is achieved.
All the Lagrangian constraints can be obtained from the Hamiltonian ones
using the $K$-operator in the following way: at each level of the Lagrangian
constraint algorithm, every Lagrangian constraint which is projectable by 
the Legendre map is associated with a Hamiltonian one
of the preceding level of the Hamiltonian constraint algorithm, which is
{\sl first-class} with respect to the {\sl primary} constraints,
whereas the non-projectable Lagrangian constraints are associated
with the {\sl second-class} Hamiltonian ones (see \cite{BGPR-eblhf}).
\item
Noether's theorem is proved and
the relation between the generators of gauge and ``rigid'' symmetries
in the Lagrangian and Hamiltonian formalisms is studied.
Thus, each Lagrangian Noether infinitessimal symmetry can be obtained from a Hamiltonian
generator of symmetries, which is a conserved quantity
(see \cite{FP-90}, \cite{GaP-00}, \cite{GP-92b}, \cite{GP-00b}, 
\cite{GP-00}, \cite{GR-93}).
\item
This operator has been applied to study Lagrangian systems whose Legendre map
has {\sl generic singularities}; that is, it degenerates on a hypersurface
(see \cite{PV-00}, \cite{PV-00b}).
\eit

Most of these results have also been generalized for 
higher-order Lagrangian systems
\cite{CL-92}, \cite{GP-95}, \cite{GPR-hols}, \cite{GPR-hoc}, and
for the case of more general types of singular differential equations on manifolds
(implicit systems of equations) \cite{GP-ggf}.
Finally, although a covariant description of this operator was not available,
it has also been used to study several characteristics of some physical models
in field theory, namely the {\sl bosonic string}
\cite{BGGP-89}, \cite{BGP-86}, \cite{GP-92b}.

Our aim is to generalize
the definition, properties and some of the applications of this operator
for field theories (specifically, the non-regular ones)
in order to describe the relationship between
the Lagrangian and Hamiltonian formalisms.
In particular, in this first work
we will study how to obtain the solutions of Lagrangian and Hamiltonian field equations
by means of this operator, and the relation between them,
while the applications concerning constraints and symmetries
will be the subject of further research.

 We follow the procedure outlined in \cite{GP-K} and \cite{CFM-95},
 which is based on the concept of {\sl section along a map} \cite{Pr-81},
 of which this operator is a particular case.
 The properties of these sections and their applications in mechanics
 have been extensively analyzed in different situations \cite{Ca-96},
 \cite{CLM-89}, \cite{CLM-91}, \cite{CMS-92}, \cite{CMS-93},
 \cite{KS-80}, \cite{PT-87}.
 Our standpoint is the
 {\sl multisymplectic jet bundle formulation} of  Lagrangian and Hamiltonian
 field theories. The essential geometric objects to be dealt with are transversal
{\sl distributions} in the corresponding bundles, which we study from
three equivalent points of view: {\sl connections}, {\sl jet fields}
and (classes of) transversal {\sl multivector fields}.
The first two are extensively used in the
standard bibliography (see, for instance, \cite{Sa-89}),
while the third is introduced as a more convenient algebraic tool
for certain applications.
Furthermore, each formulation has particular characteristics which
make easier to prove the results of the work.
  
The organization of the paper is as follows:
In Section \ref{pc}, we review first the definition and
 the main properties of the evolution operator $K$
 for autonomous mechanics. Secondly, we state the main characteristics
 of multivector fields and their relation with jet fields
 and connections in jet bundles. Then we review the Lagrangian
 and Hamiltonian multisymplectic formalisms of field theories.
Section \ref{mfo} is devoted to a study of the concept and properties
of multivector fields, jet fields and connections along maps,
 in the context of the jet bundle description of field theories.
Next, the extended and restricted covariant field operators are defined
in three equivalent ways, and their existence and non-uniqueness is proved.
Finally, in Section \ref{pmfo}, some properties of these operators are studied;
namely, how they can be used to obtain the solutions of
the Lagrangian and Hamiltonian field equations both for regular and
singular theories (on the submanifolds where they exist),
and hence the equivalence between the solutions of field
equations in the Lagrangian and
Hamiltonian formalisms are obtained in a straightforward way.

 Throughout this paper 
 $\pi\colon E\to M$ will be a fiber bundle ($\dim\, M=m$, $\dim\, E=N+m$),
 where $M$ is an oriented manifold with volume form $\omega\in\df^m(M)$.
 $\pi^1\colon J^1E\to E$ is the
 jet bundle of local sections of $\pi$, and
 $\bar\pi^1=\pi\circ\pi^1\colon J^1E \longrightarrow M$
 gives another fiber bundle structure.
 $(x^\alpha,y^A,v^A_\alpha)$ will denote natural local systems
 of coordinates in $J^1E$, adapted to the bundle $E\to M$
 ($\alpha = 1,\ldots,m$; $A= 1,\ldots,N$), and such that
 $\omega=\d x^1\wedge\ldots\wedge\d x^m\equiv\d^mx$.
 Manifolds are real, paracompact,
 connected and $C^\infty$. Maps are $C^\infty$. Sum over crossed repeated
 indices is understood.

\section{Preliminary considerations}
\protect\label{pc}

\subsection{The evolution operator $K$ in (autonomous) mechanics}
\protect\label{eok}

(See \cite{GP-K} for details).

 Let $\sigma\colon F\to B$ be a fiber bundle, and $\Phi\colon A\to B$
 a differentiable map (we assume that $\Phi(A)$ is a submanifold of $B$).
 A {\sl section along} $\Phi$ is a map ${\cal T}\colon A\to F$
 such that $\sigma\circ{\cal T} =\Phi$. So we have
$$
\begin{array}{ccc}
 & & F
\\
 & 
\begin{picture}(30,30)(0,0)
\put(8,19){\mbox{${\cal T}$}}
\put(0,0){\vector(1,1){30}}
\end{picture}
&
\begin{picture}(10,30)(0,0)
\put(8,12){\mbox{$\sigma$}}
\put(3,30){\vector(0,-1){30}}
\end{picture}
\\
A &
\begin{picture}(30,10)(0,0)
\put(13,6){\mbox{$\Phi$}}
\put(0,3){\vector(1,0){40}}
\end{picture}
& B
\end{array}
$$
In particular, if $F$ is either
$\Lambda^m\Tan B$, $\Lambda^k\Tan^*B$
or $\stackrel{k}{\otimes}\Tan^*B\stackrel{m}{\otimes}\Tan B$,
the sections are called {\sl $m$-vector fields, $k$-forms,} or
{\sl $(k,m)$-tensor fields along} $\Phi$, and the sets of these elements
are denoted by $\vf^m(A,\Phi)$, $\df^k(A,\Phi)$,
and $T^k_m(A,\Phi)$, respectively.
Obviously, every section $s\colon B\to F$
 of the projection $\sigma$ defines a section along $\Phi$ by
${\cal T}=s\circ\Phi$. 

Contractions between tensor fields along maps are defined in a natural way.
In this paper we will only use the following:
if ${\cal T}\in T^k_m(A,\Phi)$ and ${\cal T}'\in T^n_r(A,\Phi)$, then
$$
[\inn({\cal T}){\cal T}'](p):=\inn({\cal T}(p))[{\cal T}'(p)]
\quad , \quad
\mbox{\rm for every $p\in A$}
$$
so that, if $n\geq m$ then
 $\inn({\cal T}){\cal T}'\in T^{k+n-m}_r(A,\Phi)$, and
if $m=n,\, k=r=0$, then $\inn({\cal T}){\cal T}'\in\Cinfty(A)$.
Of course, if $n<m$, then $\inn({\cal T}){\cal T}'=0$.

Let $(\Tan Q,\Omega_\Lag,{\rm E}_\Lag)$ be a Lagrangian system,
${\cal F}\Lag\colon\Tan Q\to\Tan^*Q$ the Legendre map,
$\Omega\in\df^2(\Tan^*Q)$ the canonical form,
and the canonical projections
$\pi_Q\colon\Tan^*Q\to Q$, $\tau_{\Tan^*Q}\colon\Tan\Tan^*Q\to\Tan^*Q$.

The {\sl evolution operator} $K$ associated
with the Lagrangian system $(\Tan Q,\Omega_\Lag,{\rm E}_\Lag)$ is a map
$K\colon \Tan Q\longrightarrow\Tan\Tan^*Q$
satisfying the following conditions:
\ben
\item
 ({\sl Structural condition}):
 $K$ is a vector field along ${\cal F}\Lag$,
$$
\tau_{\Tan^*Q}\circ K={\cal F}\Lag
$$
\item
({\sl Dynamical condition}):
${\cal F}\Lag^*[\inn(K)(\Omega\circ{\cal F}\Lag)]=\d{\rm E}_\Lag$.
\item
({\sl Second-order condition}):
 $\Tan\pi_Q\circ K={\rm Id}_{\Tan Q}$.
\een
The existence and uniqueness of this operator can be proved, and its
local expression (using natural coordinates in $\Tan Q$ and $\Tan^*Q$) is
$$
K=v^A\left(\derpar{}{q^A}\circ{\cal F}\Lag\right)+
\derpar{\Lag}{q^A}\left(\derpar{}{p^A}\circ{\cal F}\Lag\right)
$$

By definition, $\varphi\colon\Real\to\Tan Q$ is an integral curve of $K$ if
\beq
\Tan{\cal F}\Lag\circ\dot\varphi=K\circ\varphi
\label{int00}
\eeq
so we have the diagram
$$
\begin{array}{ccccc}
 & & \Tan\Tan Q &
\begin{picture}(55,10)(0,0)
\put(12,6){\mbox{$\Tan{\cal F}\Lag$}}
\put(0,3){\vector(1,0){55}}
\end{picture} &
\Tan\Tan^*Q
\\
&
\begin{picture}(55,35)(0,0)
\put(8,18){\mbox{$\dot\varphi$}}
\put(0,0){\vector(3,2){55}}
\end{picture}
&
\begin{picture}(15,35)(0,0)
\put(13,15){\mbox{$\tau_{\Tan Q}$}}
\put(10,35){\vector(0,-1){35}}
\end{picture}
 &
\begin{picture}(55,35)(0,0)
\put(45,16){\mbox{$K$}}
\put(0,0){\vector(3,2){55}}
\end{picture}
 &
\begin{picture}(15,35)(0,0)
\put(13,15){\mbox{$\tau_{\Tan^*Q}$}}
\put(10,35){\vector(0,-1){35}}
\end{picture}
\\
\Real &
\begin{picture}(55,10)(0,0)
\put(21,6){\mbox{$\varphi$}}
\put(0,3){\vector(1,0){55}}
\end{picture}
& \Tan Q &
\begin{picture}(55,10)(0,0)
\put(20,6){\mbox{${\cal F}\Lag$}}
\put(0,3){\vector(1,0){60}}
\end{picture}
& \Tan^*Q
\end{array}
$$
Moreover, $\varphi=\dot\phi$, for $\phi\colon\Real\to Q$ 
(that is, $\varphi$ is holonomic).

The main properties of this operator are the following:
\bit
\item
If there exists an {\sl Euler-Lagrange} vector field $X_\Lag\in\vf(\Tan Q)$
for $(\Tan Q,\Omega_\Lag,{\rm E}_\Lag)$
(that is, a holonomic vector field verifying that
$\inn(X_\Lag)\Omega_\Lag=\d{\rm E}_\Lag$), then
$\varphi\colon\Real\to\Tan Q$
is an integral curve of $X_\Lag$ if, and only if,
it is an integral curve of $K$; that is, relation
(\ref{int00}) holds.

As a direct consequence of this fact,
the relation between $K$ and $X_\Lag$ is
\beq
\Tan{\cal F}\Lag\circ X_\Lag=K
\label{int000}
\eeq
In general, if the dynamical system is not regular,
Euler-Lagrange vector fields exist only
on a submanifold $S\hookrightarrow\Tan Q$.
\item
If there exists a {\sl Hamilton-Dirac} vector field $X_H\in\vf(\Tan^*Q)$
associated with the Lagrangian system $(\Tan Q,\Omega_\Lag,{\rm E}_\Lag)$
(that is, a vector field solution of the {\sl Hamilton-Dirac equations}
in the Hamiltonian formalism), then $\psi\colon\Real\to\Tan^*Q$
is an integral curve of $X_H$ if, and only if,
\beq
\dot\psi=K\circ\Tan\pi_Q\circ\dot\psi
\label{aux00}
\eeq
So we have the diagram
$$
\begin{array}{ccccc}
 & &
\Tan\Tan^*Q
& &
\\
 &
\begin{picture}(60,60)(0,0)
\put(7,45){\mbox{$\Tan\pi_Q$}}
\put(0,-2){\vector(1,1){60}}
\put(45,25){\mbox{$K$}}
\put(60,64){\vector(-1,-1){60}}
\end{picture}
 &
\begin{picture}(10,60)(0,0)
\put(8,25){\mbox{$\tau_{\Tan^*Q}$}}
\put(5,60){\vector(0,-1){60}}
\end{picture}
&
\begin{picture}(60,60)(0,0)
\put(36,35){\mbox{$\dot\psi$}}
\put(60,0){\vector(-1,1){60}}
\end{picture}
 &
\\
 \Tan Q &
\begin{picture}(60,10)(0,0)
\put(20,8){\mbox{${\cal F}\Lag$}}
\put(0,5){\vector(1,0){60}}
\end{picture}
 & \Tan^*Q &
\begin{picture}(60,10)(0,0)
\put(25,9){\mbox{$\psi$}}
\put(60,5){\vector(-1,0){60}}
\end{picture}
 &
\Real
\end{array}
$$
As a consequence, the relation between $K$ and $X_H$ is
\beq
 X_H\circ{\cal F}\Lag=K
\label{aux000}
\eeq
In general, if the dynamical system is not regular,
Hamilton-Dirac vector fields exist only
on a submanifold $P\hookrightarrow\Tan^*Q$.
\item
If $\xi\in\Cinfty(\Tan^*Q)$ is a Hamiltonian constraint,
then $\inn(K)(\d\xi\circ{\cal F}\Lag)$ is a Lagrangian constraint.
\eit

Relations (\ref{int00}), (\ref{int000}), (\ref{aux00}) and (\ref{aux000})
show how the Lagrangian and Hamiltonian descriptions can be unified by
means of the evolution operator $K$.

\subsection{Multivector fields, jet fields and connections in jet bundles}
\protect\label{mvfdm}

(See \cite{EMR-96} and \cite{EMR-98} for the proofs and other details
 of the following assertions).

Let $E$ be a $n$-dimensional differentiable manifold.
Sections of $\Lambda^m(\Tan E)$ are called
{\sl multivector fields} in $E$, or more precisely,
$m$-{\sl vector fields} in $E$
(they are contravariant skew-symmetric tensors of order $m$ in $E$).
 We will denote by $\vf^m (E)$ the set of $m$-vector fields in $E$.
 $Y\in\vf^m(E)$ is said to be {\sl locally decomposable} if,
for every $p\in E$, there exists an open neighbourhood $U_p\subset E$
and $Y_1,\ldots ,Y_m\in\vf (U_p)$ such that
$Y\feble{U_p}Y_1\wedge\ldots\wedge Y_m$.
 {\sl Contraction} of multivector fields and tensor fields in $E$
 is the usual one.

We can define the following equivalence relation:
if $Y,Y'\in\vf^m(E)$ are non-vanishing $m$-vector fields,
and $U\subseteq E$ is a connected open set,
then $Y\feblesim{U}Y'$ if there exists a
non-vanishing function $f\in\Cinfty (U)$ such that
$Y'\feble{U}fY$. Equivalence classes
will be denoted by $\{ Y\}_U$.
There is a one-to-one correspondence between the set of $m$-dimensional
orientable distributions $D$ in $\Tan E$ and the set of the
equivalence classes $\{ Y\}_E$ of non-vanishing, locally decomposable
$m$-vector fields in $E$.
If $Y\in\vf^m(E)$ is non-vanishing and locally decomposable,
 the distribution associated
with the class $\{ Y\}_U$ is denoted ${\cal D}_U(Y)$
(If $U=E$ we write ${\cal D}(Y)$).
A non-vanishing, locally decomposable
$m$-vector field $Y\in\vf^m(E)$ is said to be {\sl integrable}
(resp. {\sl involutive}) if 
 its associated distribution ${\cal D}_U(Y)$ is integrable
(resp. involutive).
Of course, if $Y\in\vf^m(E)$ is integrable (resp. involutive),
then so is every $m$-vector field in its equivalence class $\{ Y\}$,
and all of them have the same integral manifolds.
Moreover, the {\sl Frobenius' theorem} allows us to say that
a non-vanishing and locally decomposable $m$-vector field is integrable 
 if, and only if, it is involutive. 

Let us consider the following situation:
if $\pi\colon E\to M$ is a fiber bundle ($\dim\, M=m$),
we are concerned with the case where the integral manifolds of
integrable $m$-vector fields in $E$ are sections of $\pi$.
Thus, $Y\in\vf^m(E)$ is said to be {\sl $\pi$-transverse}
if, at every point $y\in E$,
$(\inn (Y)(\pi^*\omega))_y\not= 0$, where $\omega\in\df^m(M)$ is
the volume form in $M$. Hence, ${\cal D}(Y)$ is a complementary of the
vertical subbundle, and so there exists a unique connection whose horizontal
subbundle is just ${\cal D}(Y)$.
If $Y\in\vf^m(E)$ is integrable, it is $\pi$-transverse if, and only if,
its integral manifolds are local sections of $\pi\colon E\to M$.
In this case, if $\phi\colon U\subset M\to E$
is a local section with $\phi (x)=y$ and $\phi (U)$ is
the integral manifold of $Y$ through $y$,
then $\Tan_y({\rm Im}\,\phi)$ is ${\cal D}_y(Y)$.
Integral sections $\phi$ of $Y$ can be characterized
by the commutativity of the diagram
$$
\begin{array}{ccc}
\Lambda^m\Tan M &
\begin{picture}(55,10)(0,0)
\put(15,6){\mbox{$\Lambda^m\Tan\phi$}}
\put(0,3){\vector(1,0){55}}
\end{picture} &
\Lambda^m\Tan E
\\
\begin{picture}(15,35)(0,0)
\put(-12,15){\mbox{$\sigma_M$}}
\put(8,35){\vector(0,-1){35}}
\end{picture}
 & &
\begin{picture}(15,35)(0,0)
\put(13,15){\mbox{$fY$}}
\put(8,0){\vector(0,1){35}}
\end{picture}
\\
M &
\begin{picture}(55,10)(0,0)
\put(22,6){\mbox{$\phi$}}
\put(0,3){\vector(1,0){55}}
\end{picture}
& E
\end{array}
$$
that is, by the condition
\beq
\Lambda^m\Tan\phi=fY\circ\phi\circ\sigma_M
\label{int0}
\eeq
where $\sigma_M$ denotes the natural projection and 
$f\in\Cinfty (E)$ is a non-vanishing function
(observe that we are really characterizing the entire class
$\{ Y\}$ of integrable $m$-vector fields).

From the above comments we conclude that
classes of locally decomposable and $\pi$-transverse $m$-vector fields
$\{ Y\}\subset\vf^m(E)$ are in one-to-one correspondence with
orientable Ehresmann connection forms $\nabla$ in $\pi\colon E\to M$
(orientable in the sense that their associated {\sl horizontal distribution}
is orientable), and hence with orientable jet fields $\Psi\colon E\to J^1E$ \cite{Sa-89}.
Observe that this correspondence is
characterized by the fact that the horizontal subbundle
associated with $\Psi$ (and $\nabla$) coincides with ${\cal D}(Y)$.
Furthermore, the orientable jet field $\Psi$
(and the connection form $\nabla$)
is integrable if, and only if, so is $Y$, for every $Y\in\{ Y\}$.

Next we are going to make explicit the above correspondence.
Given the bundle $\pi\colon E\to M$,
denote by $\{\Lambda^m\Tan E\}$ the projective bundle
associated with $\Lambda^m\Tan E$. Let
$\rho_E\colon\Lambda^m\Tan E\to\{\Lambda^m\Tan E\}$ and
$\{\sigma_E\}\colon\{\Lambda^m\Tan E\}\to E$ be the natural projections,
and denote by $\{\Lambda^m\Tan E\}_y$ the fiber at $y\in E$.
We can define a map
$$
\Upsilon_E\colon J^1E\to\{\Lambda^m\Tan E\}
$$
as follows: for every $\bar y\in J^1E$ with
 $\bar y\stackrel{\pi^1}{\to}y\stackrel{\pi}{\mapsto}x$,
if $\phi\colon M\to E$ is a representative
of $\bar y$, then $\Upsilon_E$ maps $\bar y=j^1\phi(x)$ onto the projective
class of $m$-vectors associated with the $m$-dimensional subspace
${\rm Im}\,\Tan_x\phi$.
An element of $\{\Lambda^m\Tan E\}_y$ belongs to ${\rm Im}\,\Upsilon_E$
if it is a class of $\pi$-transverse $m$-vectors at $y\in E$,
such that it has a representative which is decomposable.
Denoting by $D_y^m$ the set of those classes, we have that
${\rm Im}\,\Upsilon_E=
\bigcup_{y\in E}D^m_y\equiv D^m\Tan E$.
The map $\Upsilon_E$ is
injective, and hence it is bijective onto its image $D^m\Tan E$.
Then, there exists its inverse $\Upsilon_E^{-1}$
(on $D^m\Tan E$), which acts as follows: for every
$(y,\{ u_1\wedge\ldots\wedge u_m\})\in D^m\Tan E$ with  $\pi(y)=x$,
let $\phi\colon M\to E$ be a local section such that
$\Tan_x\phi[\Tan_y\pi(u_i)]=u_i$, then
$\Upsilon_E^{-1}(y,\{ u_1\wedge\ldots\wedge u_m\})=j^1\phi$.

Otherwise, if $\{ Y\}\in D^m\Tan E$, there exists
a unique connection form $\nabla$ such that $\{ Y\}$ is the image in
$D^m\Tan E$ of its horizontal distribution. Let
$$
\Upsilon'_E\colon D^m\Tan E\to\pi^*\Tan^*M\otimes_E\Tan E
$$
be the map that associates to every element of $D^m\Tan E$ the corresponding
connection form (observe that, given a horizontal distribution
$H\subset\Tan E$, there is a unique map 
$\nabla\colon E\to\pi^*\Tan^*M\otimes_E\Tan E$
such that $\nabla(y)(\Tan_{\pi(y)}M)=H_y$.
Like $\Upsilon_E$, the map $\Upsilon_E'$ is injective, and hence
it is bijective onto its image, which is just the set of Ehresmann connections
in $\pi\colon E\to M$.

So we have the diagram
\bea
\begin{array}{ccc}
& & \pi^*\Tan^*M\otimes_E\Tan E \\
& &
\begin{picture}(15,50)(0,0)
\put(-13,19){\mbox{$\Upsilon'_E$}}
\put(5,0){\vector(0,1){50}}
\end{picture}
\\
\{\Lambda^m\Tan E\} & \supset &
 D^m\Tan E
\\
\begin{picture}(15,50)(0,0)
\put(-10,19){\mbox{$\rho_E$}}
\put(8,0){\vector(0,1){50}}
\end{picture}
 & &
\begin{picture}(15,50)(0,0)
\put(-13,19){\mbox{$\Upsilon_E$}}
\put(5,0){\vector(0,1){50}}
\put(13,19){\mbox{$\Upsilon_E^{-1}$}}
\put(10,50){\vector(0,-1){50}}
\end{picture}
\\
\Lambda^m\Tan E & & J^1E
\end{array}
\label{diagpro0}
\eea
Therefore, from a class
$\{ Y\}\colon E\to D^m\Tan E$ we obtain
$\Psi=\Upsilon_E^{-1}\circ \{ Y\}$,
and conversely, from $\Psi$ we construct
$\{ Y\}=\Upsilon_E\circ\Psi$.
In the same way, from the class
$\{ Y\}\colon E\to D^m\Tan E$ we obtain
$\nabla=\Upsilon_E'\circ \{ Y\}$,
and conversely, from $\nabla$ we construct
$\{ Y\}={\Upsilon_E'}^{-1}\circ\nabla$.

As an evident consequence of the existence of the bijections
$\Upsilon_E\colon J^1E\to D^m\Tan E$ and 
$\Upsilon_E'$ from $D^m\Tan E$ onto its image,
$D^m\Tan E$ inherits the affine structure over $\pi^*\Tan^*M\otimes_E{\rm V}(\pi)$
common to $J^1E$ and the set of Ehresmann connections in $\pi\colon E\to M$.

We want to characterize the integrable $m$-vector fields in $J^1E$
whose integral manifolds are canonical prolongations of sections of $\pi$.
Let $\{ X\}\colon J^1E\to D^m\Tan J^1E\subset\{\Lambda^m\Tan J^1E\}$
be a class of non-vanishing, locally decomposable
and $\bar\pi^1$-transverse $m$-vector fields in $J^1E$,
let ${\mit \Psi}\colon J^1E\to J^1J^1E$ be its associated jet field,
and $\nabla\colon J^1E\to\bar\pi^{1*}\Tan M\otimes_{J^1E}\Tan J^1E$
its associated connection form.
Now diagram (\ref{diagpro0}) can be completed as follows
\bea
\begin{array}{ccccccc}
& & \pi^*\Tan^*M\otimes_E\Tan E &
\begin{picture}(45,20)(0,0)
\put(8,9){\mbox{$\tau\otimes\Tan\pi^1$}}
\put(45,3){\vector(-1,0){45}}
\end{picture}
& \bar\pi^{1*}\Tan^*M\otimes_{J^1E}\Tan J^1E & &
\\
& &
\begin{picture}(15,50)(0,0)
\put(-13,19){\mbox{$\Upsilon'_E$}}
\put(5,0){\vector(0,1){50}}
\end{picture}
 & &
\begin{picture}(15,50)(0,0)
\put(10,0){\vector(0,1){50}}
\put(13,19){\mbox{$\Upsilon_{J^1E}'$}}
\end{picture}
& &
\\
\{\Lambda^m\Tan E\} & \supset &
 D^m\Tan E
&
\begin{picture}(45,20)(0,0)
\put(2,9){\mbox{$\{\Lambda^m\Tan\pi^1\}$}}
\put(45,3){\vector(-1,0){45}}
\end{picture}
&
D^m\Tan J^1E & \subset &
 \{\Lambda^m\Tan J^1E\}
\\
\begin{picture}(15,50)(0,0)
\put(-10,19){\mbox{$\rho_E$}}
\put(8,0){\vector(0,1){50}}
\end{picture}
 & &
\begin{picture}(15,50)(0,0)
\put(-13,19){\mbox{$\Upsilon_E$}}
\put(5,0){\vector(0,1){50}}
\put(13,19){\mbox{$\Upsilon_E^{-1}$}}
\put(10,50){\vector(0,-1){50}}
\end{picture}
 & &
\begin{picture}(15,50)(0,0)
\put(-23,19){\mbox{$\Upsilon_{J^1E}$}}
\put(5,0){\vector(0,1){50}}
\put(13,19){\mbox{$\Upsilon_{J^1E}^{-1}$}}
\put(10,50){\vector(0,-1){50}}
\end{picture}
& & 
\begin{picture}(15,50)(0,0)
\put(8,0){\vector(0,1){50}}
\put(11,19){\mbox{$\rho_{J^1E}$}}
\end{picture}
\\
\Lambda^m\Tan E & & J^1E
 &
\begin{picture}(45,10)(0,0)
\put(13,9){\mbox{$j^1\pi^1$}}
\put(45,3){\vector(-1,0){45}}
\end{picture}
&
 J^1J^1E & & \Lambda^m\Tan J^1E
\end{array}
\label{diagpro1}
\eea
where the natural projection
$\tau\otimes\Tan\pi^1\colon
\bar\pi^{1*}\Tan^*M\otimes_{J^1E}\Tan J^1E\to\pi^*\Tan^*M\otimes_E\Tan E$
acts in the following way: if 
$[\bar y,\zeta\otimes\bar v]\in\bar\pi^{1*}\Tan^*M\otimes_{J^1E}\Tan J^1E$,
with $\bar y\in J^1E$, $\zeta\in\Tan_{\bar\pi^1(\bar y)}^*M$, 
$\bar v\in\Tan_{\bar y}J^1E$,
then
\beq
(\tau\otimes\Tan\pi^1)[\bar y,\zeta\otimes\bar v]:=
[\pi^1(\bar y),\zeta\otimes\Tan_{\bar y}\pi^1(\bar v)]
\label{mvfcon0}
\eeq
Then, bearing in mind the commutativity of this diagram, we have that
${\mit\Psi}=\Upsilon_{J^1E}^{-1}\circ\{ X\}$,
and that $\nabla=\Upsilon_{J^1E}'\circ \{ X\}$, so
$$
\Upsilon_E^{-1}\circ{\Upsilon_E'}^{-1}\circ\tau\otimes\Tan\pi^1\circ\nabla=
\Upsilon_E^{-1}\circ\{\Lambda^m\Tan\pi^1\}\circ\{ X\}=
j^1\pi^1\circ{\mit\Psi}
$$
and, if $X\colon J^1E\to\Lambda^m\Tan J^1E$ is
a representative of the class $\{ X\}$, denoting
 $\varrho_E:=\Upsilon_E^{-1}\circ\{\Lambda^m\Tan\pi^1\}\circ\rho_{J^1E}$,
 from the last equality we obtain that
$$
\Upsilon_E^{-1}\circ{\Upsilon_E'}^{-1}\circ\tau\otimes\Tan\pi^1\circ\nabla=
\varrho_E\circ X=j^1\pi^1\circ{\mit\Psi}
$$

\begin{definition}
The jet field ${\mit\Psi}$, its associated connection form $\nabla$
in $\bar\pi^1\colon J^1E\to M$,
and their associated class $\{ X\}$ are said to be:
\ben
\item
{\rm Semi-holonomic}
(or a {\rm Second Order Partial Differential Equation}), if
$$
\Upsilon_E^{-1}\circ{\Upsilon_E'}^{-1}\circ\tau\otimes\Tan\pi^1\circ\nabla=
\Upsilon_E^{-1}\circ\{\Lambda^m\Tan\pi^1\}\circ\{ X\}=
j^1\pi^1\circ{\mit\Psi}={\rm Id}_{J^1E}
$$
If $X\in\{ X\}$ is a representative of this class, then it is
a semi-holonomic $m$-vector field, and
the above condition leads to
$$
\varrho_E\circ X={\rm Id}_{J^1E}
$$
\item
{\rm Holonomic} if
 they are integrable and their integral sections
 $\varphi\colon M\to J^1E$ are holonomic (that is,
$\varphi=j^1\phi$, for some section $\varphi\colon M\to E$).
\een
\end{definition}

Then, it can be proved that the class $\{ X\}$,
and its associated jet field ${\mit\Psi}$ and connection form $\nabla$
are holonomic if, and only if, they are integrable and semi-holonomic.

In a natural chart in $J^1E$, the local expressions of these elements are
\bea
 X &=&
 \bigwedge_{\alpha=1}^m f
\left(\derpar{}{x^\alpha}+F_\alpha^A\derpar{}{y^A}+
 G_{\alpha\nu}^A\derpar{}{v_\nu^A}\right)
\nonumber \\
{\mit\Psi}&=&
(x^\alpha,y^A,v^A_\alpha,F^A_\alpha,G^A_{\alpha\nu})
\nonumber \\
\nabla &=& \d x^\alpha\otimes
\left(\derpar{}{x^\alpha}+F_\alpha^A\derpar{}{y^A}+
 G_{\alpha\nu}^A\derpar{}{v_\nu^A}\right)
 \label{sopde}
 \eea
where $f\in\Cinfty(J^1E)$ is an arbitrary non-vanishing function.
 A representative of the class $\{ X\}$ can be selected by the condition
 $\inn (X)(\bar\pi^{1*}\omega)=1$, which leads to
 $f=1$ in the above local expression.
 We will adopt this particular choice in the sequel.

Now, if these elements are integrable, and
$\varphi(x)=(x^\alpha,y^A=\varphi^A(x),v^A_\alpha=\varphi^A_\alpha(x))$
is an integral section, then the components of $\varphi$ are solution of the system 
of partial differential equations
 $$
 F_\alpha^A(x,\varphi^B(x),\varphi^B_\eta(x))=
 \derpar{\varphi^A}{x^\alpha}
\ ,\quad
 G_{\alpha\nu}^A(x,\varphi^B(x),\varphi^B_\eta(x))=
 \derpar{\varphi^A_\alpha}{x^\nu}
\ ;\quad
(A,B=1\ldots N;\ \eta,\alpha,\nu=1\ldots m)
 $$
If these elements are semi-holonomic, their local expressions
are the same as in (\ref{sopde}) with $F^A_\alpha=v^A_\alpha$.
In addition, if they are integrable, as their integral sections are holonomic,
given a section $\phi(x)=(x^\alpha,y^A=\varphi^A(x))$, its jet prolongation
\dst j^1\phi(x)=\left( x^\alpha,y^A=\varphi^A(x),
v^A_\alpha=\derpar{\varphi^A}{x^\alpha}(x)\right)\) 
is an integral section iff the components of $\phi$ are solution of the
second-order system of partial differential equations
 \beq
 G_{\alpha\nu}^A\left(x,\varphi^B,\derpar{\varphi^B}{x^\eta}\right) =
 \frac{\partial^2\phi^A}{\partial x^\nu\partial x^\alpha}
\quad ;\quad
(A,B=1,\ldots ,N\ ;\ \eta,\alpha,\nu=1,\ldots ,m)
 \label{sopdeqs}
 \eeq
(See \cite{EMR-96} and \cite{EMR-98} for details).

\subsection{Lagrangian formalism for classical field theories}
\protect\label{lf}

(See, for instance, \cite{BSF-88}, \cite{EMR-96}, \cite{EMR-98}, 
 \cite{GMS-97}, \cite{LMM-96}, \cite{Sd-95}, \cite{Sa-89}, for details).

A {\sl classical field theory} is described by its
{\sl configuration fiber bundle} $\pi\colon E\to M$;
and a {\sl Lagrangian density} which is
a $\bar\pi^1$-semibasic $m$-form on $J^1E$.
A Lagrangian density is usually written as
$\Lag =\lag \bar\pi^{1^*}\omega$, where $\lag\in\Cinfty (J^1E)$
is the {\sl Lagrangian function} associated with $\Lag$ and $\omega$.
The {\sl Poincar\'e-Cartan $m$ and $(m+1)$-forms}
associated with the Lagrangian density $\Lag$
are defined using the {\sl vertical endomorphism}
${\cal V}$ of the bundle $J^1E$
$$
\Theta_{\Lag}:=\inn({\cal V})\Lag+\Lag\in\df^{m}(J^1E)
\quad ;\quad
\Omega_{\Lag}:= -\d\Theta_{\Lag}\in\df^{m+1}(J^1E)
$$
Then a {\sl Lagrangian system} is a couple $\ls$.
We will say that the Lagrangian system is {\sl regular} if
 $\Omega_{\Lag}$ is $1$-nondegenerate.
In a natural chart in $J^1E$ we have
\beann
\Omega_{\Lag}&=&
-\frac{\partial^2\lag}{\partial v^B_\nu\partial v^A_\alpha}
\d v^B_\nu\wedge\d y^A\wedge\d^{m-1}x_\alpha 
-\frac{\partial^2\lag}{\partial y^B\partial v^A_\alpha}\d y^B\wedge
\d y^A\wedge\d^{m-1}x_\alpha +
 \\  & &
\frac{\partial^2\lag}{\partial v^B_\nu\partial v^A_\alpha}v^A_\alpha
\d v^B_\nu\wedge\d^mx  +
\left(\frac{\partial^2\lag}{\partial y^B\partial v^A_\alpha}v^A_\alpha
 -\derpar{\lag}{y^B}+
\frac{\partial^2\lag}{\partial x^\alpha\partial v^B_\alpha}
\right)\d y^B\wedge\d^mx
\eeann
(where \dst\d^{m-1}x_\alpha\equiv\inn\left(\derpar{}{x^\alpha}\right)\d^mx\) );
and the regularity condition is equivalent to
\dst det\left(\frac{\partial^2\lag}
{\partial v^A_\alpha\partial v^B_\nu}(\bar y)\right)\not= 0\) ,
for every $\bar y\in J^1E$.
We must point out that, in field theories, the notion of regularity
is not uniquely defined (for other approaches see, for instance,
\cite{Be-84}, \cite{De-77}, \cite{De-78}, \cite{Kr-87}, \cite{KS-01a}, 
\cite{KS-01b}).

The {\sl Lagrangian problem} associated with a Lagrangian system
 $\ls$ consists in finding sections $\phi\in\Gamma(M,E)$
($\Gamma(M,E)$ denotes the set of sections of $\pi$),
which are characterized by the condition
 $$
 (j^1\phi)^*\inn (X)\Omega_\Lag=0 \quad ,\quad
 \mbox{\rm for every $X\in\vf (J^1E)$}
 $$
 In natural coordinates, if $\phi(x)=(x^\alpha,\phi^A(x))$,
 this condition is equivalent to demanding that
 $\phi$ satisfy the {\sl Euler-Lagrange equations}
 \beq
 \derpar{\lag}{y^A}\Big\vert_{j^1\phi}-
\derpar{}{x^\alpha}\derpar{\lag}{v_\alpha^A}\Big\vert_{j^1\phi} = 0
 \quad , \quad \mbox{\rm (for $A=1,\ldots ,N$)}
 \label{eleqs}
 \eeq

 The problem of finding these sections
 can be formulated equivalently as follows:
 finding a distribution $D$ of $\Tan (J^1E)$ such that
 it is integrable (that is, {\sl involutive\/}),
 $m$-dimensional, $\bar\pi^1$-transverse, and
 the integral manifolds of $D$ are the sections solution
 of the above equations.
 This is equivalent to stating that the sections solution of
the Lagrangian problem are the integral sections of a class of
holonomic $m$-vector fields $\{ X_{\Lag}\}\subset\vf^m(J^1E)$, such that
 $$
 \inn (X_{\Lag})\Omega_{\Lag}=0
 \quad , \quad
 \mbox{\rm for every $X_\Lag\in\{ X_{\Lag}\}$}
 $$
Taking into account the equivalence between classes of 
non-vanishing, locally decomposable and $\bar\pi^1$-transverse
$m$-vector fields with orientable jet fields and connections,
we can also state the above problem in two additional equivalent ways:
finding a holonomic connection
 $\nabla_\Lag$ in $\bar\pi^1\colon J^1E\to M$
such that
$$
 \inn(\nabla_{\Lag})\Omega_{\Lag}=(m-1)\Omega_\Lag
$$
or a holonomic jet field ${\mit\Psi}_\Lag\colon J^1E\to J^1J^1E$, such that
$$
 \inn ({\mit\Psi}_{\Lag})\Omega_{\Lag}=0
$$
(where the contraction of jet fields with differential forms
is defined in \cite{EMR-96}).
 Semi-holonomic locally decomposable $m$-vector fields,
 jet fields and connections which are solution of these equations
 are called {\sl Euler-Lagrange $m$-vector fields, jet fields}
 and {\sl connections} for $\ls$.

 Their local expressions are (\ref{sopde}) with $F^A_\alpha=v^A_\alpha$,
 and where the coefficients
 $G^A_{\alpha\nu}$ are related by the system of linear equations
 \beq
\frac{\partial^2\lag}{\partial v^A_\alpha\partial  v^B_\nu}G^A_{\alpha\nu}=
\derpar{\lag}{y^B}-\frac{\partial^2\lag}{\partial x^\nu\partial v^B_\nu}-
\frac{\partial^2\lag}{\partial y^A\partial v^B_\nu}v^A_\nu
\qquad (A,B=1,\ldots ,N)
 \label{eqsG0}
 \eeq
Therefore, if \dst j^1\phi=
\left( x^\mu,\phi^A,\derpar{\phi^A}{x^\nu}\right)\)
is an integral section of $X_\Lag$, then
\dst v^A_\alpha=\derpar{\phi^A}{x^\alpha}\) , and hence
the coefficients $G^B_{\alpha\nu}$ must satisfy equations (\ref{sopdeqs}).
As a consequence, the system (\ref{eqsG0}) is equivalent 
to the Euler-Lagrange equations (\ref{eleqs}) for the section $\phi$.

 If $\ls$ is a regular Lagrangian system, then
 the existence of classes of Euler-Lagrange $m$-vector fields for $\Lag$
(or what is equivalent, Euler-Lagrange jet fields or connections)
 is assured, and in a local system of coordinates,
 these $m$-vector fields depend on $N(m^2-1)$ arbitrary functions.
For singular Lagrangian systems,
the existence of Euler-Lagrange $m$-vector fields
 is not assured except perhaps
on some submanifold $S\hookrightarrow J^1E$, and
 the number of arbitrary functions on which they depend
is not the same as in the regular case, since it depends on the dimension of
$S$ and the rank of the Hessian matrix of $\lag$.
Furthermore, locally decomposable and $\bar\pi^1$-transverse
$m$-vector fields, solutions of the field equations
can exist (in general, on some submanifold of $J^1E$),
but none of them being semi-holonomic (at any point of this submanifold).
As in the regular case, although Euler-Lagrange $m$-vector fields exist
on some submanifold $S$, their integrability is not assured except
perhaps on another smaller submanifold $I\hookrightarrow S$ such that
the integral sections are contained in $I$.

\subsection{Hamiltonian formalism for classical field theories}
\protect\label{hf}

(See, for instance,  \cite{CCI-91}, \cite{EMR-99b},
 \cite{EMR-00}, \cite{GMS-97}, \cite{HK-01},
\cite{Ka-98}, \cite{LMM-96} for details).

For the Hamiltonian formalism of field theories,
the choice of a {\sl multimomentum phase space}
 or {\sl multimomentum bundle} is not unique. In this work we take:
 $ J^{1*}E\equiv\Lambda_1^m\Tan^*E/\Lambda^m_0\Tan^*E$,
 where $\Lambda_1^m\Tan^*E\equiv {\cal M}\pi$
 is the bundle of $m$-forms on
 $E$ vanishing by the action of two $\pi$-vertical vector fields
(sometimes it is called the {\sl extended multimomentum bundle}\/),
 and $\Lambda^m_0\Tan^*E\equiv\pi^*\Lambda^m\Tan^*M$.
 We have the natural projections
 $$
 \tau^1\colon J^{1*}E\to E \quad ,\quad
 \bar\tau^1=\pi\circ\tau^1\colon J^{1*}E\to M
 $$
 Given a system of coordinates adapted to the bundle
 $\pi\colon E\to M$, we can construct natural coordinates in
 $J^{1*}E$ and ${\cal M}\pi$, which will be denoted as
 $(x^\alpha,y^A,p^\alpha_A)$ and $(x^\alpha,y^A,p^\alpha_A,p)$, respectively
 ($\alpha = 1,\ldots,m$; $A= 1,\ldots,N$).

 Now, if $\ls$ is a Lagrangian system, we
 introduce the {\sl extended Legendre map} associated with $\Lag$,
 $\widetilde{{\cal F}\Lag}\colon J^1E\to {\cal M}\pi$,
 in the following way:
  $$
 (\widetilde{{\cal F}\Lag}\bar y))(\moment{Z}{1}{m}):=
 (\Theta_{\Lag})_{\bar y}(\moment{\bar Z}{1}{m})
 $$
 where $\moment{Z}{1}{m}\in\Tan_{\pi^1(\bar y)}E$, and
 $\moment{\bar Z}{1}{m}\in\Tan_{\bar y}J^1E$ are such that
 $\Tan_{\bar y}\pi^1\bar Z_\alpha=Z_\alpha$.
 ($\widetilde{{\cal F}\Lag}$ can also be defined as the
  ``first order  vertical Taylor approximation to
 $\lag$'' \cite{CCI-91}).
 Hence, using the natural projection
 $\mu \colon {\cal M}\pi=\Lambda_1^m\Tan^*E \to
 \Lambda_1^m\Tan^*E/\Lambda_0^m\Tan^*E=J^{1*}E$,
 we define the {\sl restricted Legendre map} associated with $\Lag$ as
 ${\cal F}\Lag :=\mu\circ\widetilde{{\cal F}\Lag}$.
 Their local expressions are
 $$
 \begin{array}{ccccccc}
 \widetilde{{\cal F}\Lag}^*x^\alpha = x^\alpha &\quad\ , \ \quad&
 \widetilde{{\cal F}\Lag}^*y^A = y^A &\quad\  , \quad&
 \widetilde{{\cal F}\Lag}^*p_A^\alpha =\derpar{\lag}{v^A_\alpha}
 &\quad\ , \quad&
 \widetilde{{\cal F}\Lag}^*p =\lag-v^A_\alpha\derpar{\lag}{v^A_\alpha}
 \\
 {\cal F}\Lag^*x^\alpha = x^\alpha &\quad\ , \ \quad&
 {\cal F}\Lag^*y^A = y^A &\quad\  , \quad&
 {\cal F}\Lag^*p_A^\alpha =\derpar{\lag}{v^A_\alpha} & &
 \end{array}
 $$
 Then, $\ls$ is a {\sl regular}
 Lagrangian system if ${\cal F}\Lag$ is a local diffeomorphism
 (this definition is equivalent to that given above).
 Elsewhere $\ls$ is a {\sl singular} Lagrangian system.
 As a particular case, $\ls$ is a {\sl hyper-regular}
 Lagrangian system if ${\cal F}\Lag$ is a global diffeomorphism.
 A singular Lagrangian system $\ls$ is {\sl almost-regular} if:
 ${\cal P}:={\cal F}\Lag (J^1E)$ is a closed submanifold of $J^{1*}E$
 (we will denote the natural imbedding by
  $\jmath_0\colon {\cal P}\hookrightarrow J^{1*}E$),
 ${\cal F}\Lag$ is a submersion onto its image, and
 for every $\bar y\in J^1E$, the fibres
 ${\cal F}\Lag^{-1}({\cal F}\Lag (\bar y))$
 are connected submanifolds of $J^1E$.
 
 In order to construct a {\sl Hamiltonian system} associated with $\ls$,
 first, recall that the multicotangent bundle
 $\Lambda^m\Tan^*E$ is endowed with canonical forms:
 ${\bf \Theta}\in\df^m(\Lambda^m\Tan^*E)$
 and the multisymplectic form
 ${\bf  \Omega}:=-\d{\bf \Theta}\in\df^{m+1}(\Lambda^m\Tan^*E)$.
 But ${\cal M}\pi\equiv\Lambda^m_1\Tan^*E$ is a subbundle of
 $\Lambda^m\Tan^*E$. Then, if
 $\lambda\colon\Lambda^m_1\Tan^*E\hookrightarrow\Lambda^m\Tan^*E$
 is the natural imbedding,
 $\Theta :=\lambda^* {\bf \Theta}$ and
 $\Omega :=-\d\Theta=\lambda^*{\bf \Omega}$
 are canonical forms in ${\cal M}\pi$, which are called
 the {\sl multimomentum Liouville} $m$ and $(m+1)$ {\sl forms}.
 Their local expressions are
 \beq
 \Theta = p_A^\alpha\d y^A\wedge\d^{m-1}x_\alpha+p\d^mx
 \quad , \quad
 \Omega = -\d p_A^\alpha\wedge\d y^A\wedge\d^{m-1}x_\alpha-\d p\wedge\d^mx
 \label{Omegah}
 \eeq
Observe that
 $\widetilde{{\cal F}\Lag}^*\Theta=\Theta_{\Lag}$,
 and $\widetilde{{\cal F}\Lag}^*\Omega=\Omega_{\Lag}$.

 Now, if $\ls$ is a hyper-regular Lagrangian system, then
 $\tilde{\cal P}:=\widetilde{{\cal F}\Lag}(J^1E)$ is a
 1-codimensional imbedded submanifold of ${\cal M}\pi$
 (we will denote the natural imbedding by
  $\tilde\jmath_0\colon\tilde{\cal P}\hookrightarrow{\cal M}\pi$),
 which is transverse to the projection $\mu$, and is diffeomorphic to
 $J^{1*}E$. This diffeomorphism is $\mu^{-1}$, when $\mu$ is
 restricted to $\tilde{\cal P}$, and also coincides with the map
 $h:=\widetilde{{\cal F}\Lag}\circ{\cal F}\Lag^{-1}$,
 when it is restricted onto its image (which is just $\tilde{\cal P}$).
 This map $h$ is called a {\sl Hamiltonian section}, and
 can be used to construct the {\sl Hamilton-Cartan} $m$ and $(m+1)$ {\sl forms}
 of $J^{1*}E$ by making
 $$
 \Theta_h=h^*\Theta\in\df^m(J^{1*}E)
\quad , \quad
\Omega_h=h^*\Omega\in\df^{m+1}(J^{1*}E)
 $$
 and the couple $\hs$ is said to be the {\sl Hamiltonian system}
 associated with the hyper-regular Lagrangian system $\ls$.
 Locally, the Hamiltonian section $h$ is specified by the
 {\sl local Hamiltonian function}
 $H=p^\alpha_A ({\cal F}\Lag^{-1})^*v_\alpha^A-({\cal F}\Lag^{-1})^*\lag$,
 that is, $h(x^\alpha,y^A,p^\alpha_A)=(x^\alpha,y^A,p^\alpha_A,-H)$.
 Then we have the local expressions
 $$
 \Theta_h = p_A^\alpha\d y^A\wedge\d^{m-1}x_\alpha -H\d^mx
 \quad , \quad
 \Omega_h = -\d p_A^\alpha\wedge\d y^A\wedge\d^{m-1}x_\alpha +
 \d H\wedge\d^mx
 $$
  Of course
 ${\cal F}\Lag^*\Theta_h=\Theta_{\Lag}$,
 and ${\cal F}\Lag^*\Omega_h=\Omega_{\Lag}$.

 The {\sl Hamiltonian problem} associated with the Hamiltonian
 system $\hs$ consists in finding
 sections $\psi\in\Gamma(M,J^{1*}E)$,
 which are characterized by the condition 
 $$
 \psi^*\inn (X)\Omega_h=0 \quad , \quad
 \mbox{\rm  for every $X\in\vf (J^{1*}E)$}
 $$
 In natural coordinates, if
 $\psi(x)=(x^\alpha,y^A(x),p^\alpha_A(x))$, this condition leads to the
 so-called {\sl Hamilton-De Donder-Weyl equations}
 $$
 \derpar{y^A}{x^\alpha}\Bigg\vert_{\psi}=
 \derpar{H}{p^\alpha_A}\Bigg\vert_{\psi}
 \quad ;\quad
 \derpar{p_A^\alpha}{x^\alpha}\Bigg\vert_{\psi}=
 -\derpar{H}{y^A}\Bigg\vert_{\psi}
 $$
 The problem of finding these sections
 can be formulated equivalently as follows:
 finding a distribution $D$ of $\Tan (J^{1*}E)$ such that
 $D$ is integrable (that is, {\sl involutive\/}),
 $m$-dimensional, $\bar\tau^1$-transverse, and
 its integral manifolds are the sections solution of the
 above equations. This is equivalent to stating that
 the sections solution of
 the Hamiltonian problem are the integral sections of a class of
 integrable and $\bar\tau^1$-transverse $m$-vector fields
 $\{ X_{\cal H}\}\subset\vf^m(J^{1*}E)$ satisfying that
 $$
 \inn (X_{\cal H})\Omega_h=0 \quad ,   \quad
 \mbox{\rm for every $X_{\cal H}\in\{ X_{\cal H}\}$}
 $$
As in the Lagrangian formalism,
we can also state the above problem in two additional equivalent ways:
finding an orientable connection $\nabla_{\cal H}$ 
in $\bar\tau^1\colon J^{1*}E\to M$ such that
$$
 \inn(\nabla_{\cal H})\Omega_h=(m-1)\Omega_h
$$
or an orientable jet field ${\mit\Psi}_{\cal H}\colon J^{1*}E\to J^1J^{1*}E$,
 such that
$$
 \inn ({\mit\Psi}_{\cal H})\Omega_h=0
$$
$\bar\tau^1$-transverse and locally decomposable $m$-vector fields,
 orientable jet fields and orientable connections
 which are solutions of these equations
 are called {\sl Hamilton-De Donder-Weyl (HDW) $m$-vector fields,
 jet fields} and {\sl connections} for $\hs$.

 Their local expressions in natural coordinates are
\beann
  X_{{\cal H}} &=& \bigwedge_{\alpha=1}^m
 f\left(\derpar{}{x^\alpha}+F_\alpha^A\derpar{}{y^A}+
 G^\eta_{A\alpha}\derpar{}{p^\eta_A}\right)
  \\
{\mit\Psi}_{{\cal H}} &=&
(x^\alpha,y^A,p_A^\alpha,;F^A_\alpha,G^\eta_{A\alpha})
 \\
\nabla_{{\cal H}} &=& \d x^\alpha\otimes
\left(\derpar{}{x^\alpha}+F_\alpha^A\derpar{}{y^A}+
 G_{A\alpha}^\nu\derpar{}{p^\nu_A}\right)
 \eeann
 where $f\in\Cinfty(J^{1*}E)$ is a non-vanishing function,
 and the coefficients $F_\alpha^A,G^\eta_{A\alpha}$ are related by the
 system of linear equations
 $$
 F^A_\alpha=\derpar{H}{p_A^\alpha}
 \quad , \quad
 G^\nu_{A\nu}=-\derpar{H}{y^A}
 $$
 Now, if $\psi(x)=(x^\alpha ,\psi^A(x),\psi^\alpha_A(x))$
 is an integral section of $X_{\cal H}$ then
 $$
 F^A_\alpha\circ \psi =\derpar{\psi^A}{x_\alpha} \quad ; \quad
 G^\alpha_{A\alpha}\circ\psi = \derpar{\psi^\alpha_A}{x^\alpha}
 $$
 which are the Hamilton-De Donder-Weyl equations for $\psi$.
 As above, a representative of the class $\{ X\}$
 can be selected by the condition
 $\inn (X)(\bar\tau^{1*}\omega)=1$, which leads to $f=1$.

 The existence of classes of HDW $m$-vector fields,
 jet fields and connections
 is assured, and in a local system of coordinates they depend on $N(m^2-1)$
 arbitrary functions.

 In an analogous way, if $\ls$ is an almost-regular Lagrangian system,
 the submanifold $\jmath_0\colon {\cal P}\hookrightarrow J^{1*}E$,
 is a fiber bundle over $E$ (and $M$). The
 corresponding projections will be denoted by
 $\tau^1_0\colon {\cal P}\to E$ and $\bar\tau^1_0\colon{\cal P}\to M$,
 satisfying that $\tau^1\circ\jmath_0=\tau^1_0$ and
 $\bar\tau^1\circ\jmath_0=\bar\tau^1_0$.
 In this case the $\mu$-transverse submanifold
 $\tilde{\cal P}\hookrightarrow{\cal M}\pi$ is diffeomorphic to
 ${\cal P}$. This diffeomorphism is denoted
 $\tilde\mu\colon\tilde{\cal P}\to{\cal P}$,
 and it is just the restriction of the projection $\mu$ to $\tilde{\cal P}$.
 Then, taking
 $\tilde h:=\tilde\mu^{-1}=\widetilde{{\cal F}\Lag}_0\circ{\cal F}\Lag_0^{-1}$,
 (where $\widetilde{{\cal F}\Lag}_0$ and ${\cal F}\Lag_0$
 are the restriction maps of $\widetilde{{\cal F}\Lag}$
 and ${\cal F}\Lag$ onto $\tilde{\cal P}$ and ${\cal P}$, respectively),
 we define the Hamilton-Cartan forms
  $$
 \Theta^0_h=(\tilde\jmath_0\circ\tilde h)^*\Theta
 \quad ; \quad
 \Omega^0_h=(\tilde\jmath_0\circ\tilde h)^*\Omega
 $$
 which verify that
 ${\cal F}\Lag_0^*\Theta^0_h=\Theta_{\Lag}$ and
 ${\cal F}\Lag_0^*\Omega^0_h=\Omega_{\Lag}$.
 Then $\hso$ is the {\sl Hamiltonian system}
 associated with the almost-regular Lagrangian system $\ls$,
 and we have the following diagram
 \beq
\begin{array}{cccc}
\begin{picture}(15,52)(0,0)
\put(0,0){\mbox{$J^1E$}}
\end{picture}
&
\begin{picture}(65,52)(0,0)
 \put(17,28){\mbox{$\widetilde{{\cal F}\Lag}_0$}}
 \put(24,7){\mbox{${\cal F}\Lag_0$}}
 \put(0,7){\vector(2,1){65}}
 \put(0,4){\vector(1,0){65}}
\end{picture}
&
\begin{picture}(90,52)(0,0)
 \put(5,0){\mbox{${\cal P}$}}
 \put(5,42){\mbox{$\tilde{\cal P}$}}
 \put(5,13){\vector(0,1){25}}
 \put(10,38){\vector(0,-1){25}}
 \put(-5,22){\mbox{$\tilde h$}}
 \put(12,22){\mbox{$\tilde\mu$}}
 \put(30,45){\vector(1,0){55}}
 \put(30,4){\vector(1,0){55}}
 \put(48,12){\mbox{$\jmath_0$}}
 \put(48,33){\mbox{$\tilde\jmath_0$}}
 \end{picture}
&
\begin{picture}(15,52)(0,0)
 \put(0,0){\mbox{$J^{1*}E$}}
 \put(0,41){\mbox{${\cal M}\pi$}}
 \put(10,38){\vector(0,-1){25}}
 \put(0,22){\mbox{$\mu$}}
\end{picture}
\\
& &
\begin{picture}(90,35)(0,0)
 \put(10,35){\vector(1,-1){35}}
 \put(5,11){\mbox{$\bar\tau^1_0$}}
 \put(90,11){\mbox{$\bar\tau^1$}}
 \put(100,35){\vector(-1,-1){35}}
\end{picture}
 &
\\
& & \quad\quad M &
 \end{array}
\label{arhs}
 \eeq

 The {\sl Hamiltonian problem} associated with the Hamiltonian
 system $\hso$ is stated as in the regular case, and
 the sections $\psi_o\in\Gamma(M,{\cal P})$ solution of
 the Hamiltonian problem are the integral sections of a class of
 integrable and $\bar\tau^1_0$-transverse $m$-vector fields
 $\{ X_{{\cal H}_o}\}\subset\vf^m({\cal P})$ satisfying that
 $$
 \inn (X_{{\cal H}_o})\Omega^0_h=0 \quad ,   \quad
 \mbox{\rm for every $X_{{\cal H}_o}\in\{ X_{{\cal H}_o}\}$}
 %\label{hameq0}
 $$
As above, this is equivalent to finding an orientable connection
 $\nabla_{{\cal H}_o}$ in $\bar\tau^1_0\colon{\cal P}\to M$ such that
$$
 \inn(\nabla_{{\cal H}_o})\Omega_h^0=(m-1)\Omega_h^0
$$
or an orientable jet field
 ${\mit\Psi}_{{\cal H}_o}\colon{\cal P}\to J^1{\cal P}$,
 such that
$$
 \inn ({\mit\Psi}_{{\cal H}_o})\Omega_h^0=0
$$
 Now, not even the existence of
 these Hamilton-De Donder-Weyl $m$-vector fields,
 jet fields and connections for $\hso$
 is assured, and an algorithmic procedure in order to obtain a submanifold
 $P$ of ${\cal P}$ where such $m$-vector fields, jet fields and connections
 exist, can be outlined.
 Of course, in general, the solution is not unique,
 but the number of arbitrary functions
 is not the same as above (it depends on the dimension of $P$).

\section{The field operators}
\protect\label{mfo}

\subsection{Sections along the Legendre maps in field theories}
\protect\label{alongfl}

 First remark that for multivector fields along maps
 the terminology introduced in Section \ref{mvfdm}
 will be applied in a natural way. Thus, for instance:
 \bit
 \item
 If ${\cal X}$ is a $m$-vector field along $\Phi$,
 it is locally decomposable if, for every $p\in A$,
 there exists an open neighbourhood $U_p\subset A$,
 and ${\cal X}_1,\ldots ,{\cal X}_m$, vector fields along $\Phi$, such that
 ${\cal X}\feble{U_p}{\cal X}_1\wedge\ldots\wedge{\cal X}_m$.
 \item
 If ${\cal X},{\cal X}'$ are non-vanishing multivector fields along $\Phi$,
 and $U\subseteq A$ is a connected open set,
 then ${\cal X}\feblesim{U}{\cal X}'$ if there exists a
 non-vanishing function $f\in\Cinfty (U)$ such that
 ${\cal X}'\feble{U}f{\cal X}$.
 \eit

Now, let $\pi\colon E\to M$ be
the configuration fiber bundle of a Lagrangian system $\ls$.
The case we will consider consists in
taking $A\equiv J^1E$, $B={\cal M}\pi$, and $\Phi=\widetilde{{\cal F}\Lag}$.

\begin{definition}
\ben
\item
A {\rm $m$-vector field along} $\widetilde{{\cal F}\Lag}$ is
a map $\tilde{\cal X}\colon J^1E\to\Lambda^m\Tan{\cal M}\pi$ such that
$$
\sigma_{{\cal M}\pi}\circ\tilde{\cal X}=\widetilde{{\cal F}\Lag}
$$
where $\sigma_{{\cal M}\pi}\colon\Lambda^m\Tan{\cal M}\pi\to{\cal M}\pi$
is the natural projection.
\item 
A {\rm jet field along} $\widetilde{{\cal F}\Lag}$ is
a map $\tilde{\cal Y}\colon J^1E\to J^1{\cal M}\pi$ such that
$$
\pi^1_{{\cal M}\pi}\circ\tilde{\cal Y}=\widetilde{{\cal F}\Lag}
$$
where $\pi^1_{{\cal M}\pi}\colon J^1{\cal M}\pi\to{\cal M}\pi$
is the natural projection.
\item
An {\rm Ehresmann connection form along} $\widetilde{{\cal F}\Lag}$ is a map 
$\tilde\nabla\colon J^1E\to
(\bar\tau^1\circ\mu)^*\Tan^*M\otimes_{{\cal M}\pi}\Tan{\cal M}\pi$
such that
$$
\kappa_{{\cal M}\pi}\circ\tilde\nabla=\widetilde{{\cal F}\Lag}
$$
where
$\kappa_{{\cal M}\pi}\colon
(\bar\tau^1\circ\mu)^*\Tan^*{\cal M}\pi\otimes_M\Tan{\cal M}\pi\to{\cal M}\pi$
is the natural projection, and satisfying that
$$
\inn(\tilde\nabla)\tilde{\mit \Xi}=\tilde{\mit \Xi}
$$
for every $(\bar\tau^1\circ\mu)$-semibasic 1-form $\tilde{\mit \Xi}$ along
$\widetilde{{\cal F}\Lag}$.
(Observe that $\tilde\nabla$ is a (1,1)-tensor field along
$\widetilde{{\cal F}\Lag}$).
\een
\end{definition}

Recall that a 1-form ${\mit \Xi}$ along $\widetilde{{\cal F}\Lag}$ is
$(\bar\tau^1\circ\mu)$-semibasic if, for every
$Z\in\vf^{{\rm V}(\bar\tau^1\circ\mu)}({\cal M}\pi)$, and $\bar y\in J^1E$,
we have that
$\inn(Z_{\widetilde{{\cal F}\Lag}(\bar y)})({\mit \Xi}(\bar y))=0$.
In particular, if $\xi\in\df^1({\cal M}\pi)$ is a
$(\bar\tau^1\circ\mu)$-semibasic form, then
${\mit \Xi}:=\xi\circ\widetilde{{\cal F}\Lag}$ is a
$(\bar\tau^1\circ\mu)$-semibasic 1-form along $\widetilde{{\cal F}\Lag}$.

So we have the diagrams
$$
\begin{array}{ccc}
 & &
\begin{picture}(10,10)(0,0)
\put(-15,0)
{\mbox{$\Lambda^m\Tan{\cal M}\pi$}}
\end{picture}
\\
 & 
\begin{picture}(30,30)(0,0)
\put(8,19){\mbox{$\tilde{\cal X}$}}
\put(0,0){\vector(1,1){30}}
\end{picture}
&
\begin{picture}(15,30)(0,0)
\put(10,12){\mbox{$\sigma_{{\cal M}\pi}$}}
\put(5,30){\vector(0,-1){30}}
\end{picture}
\\
 J^1E &
\begin{picture}(30,10)(0,0)
\put(10,6){\mbox{$\widetilde{{\cal F}\Lag}$}}
\put(-2,3){\vector(1,0){40}}
\end{picture}
& {\cal M}\pi
\\
&
\begin{picture}(40,35)(0,0)
 \put(0,35){\vector(1,-2){15}}
 \put(-5,11){\mbox{$\bar\tau^1_0$}}
 \put(38,11){\mbox{$\bar\tau^1$}}
 \put(45,35){\vector(-1,-2){15}}
\end{picture}
&
\\
& M &
\end{array}
\quad 
\begin{array}{ccc}
 & &
\begin{picture}(10,10)(0,0)
\put(-12,0)
{\mbox{$J^1{\cal M}\pi$}}
\end{picture}
\\
 & 
\begin{picture}(30,30)(0,0)
\put(8,19){\mbox{$\tilde{\cal Y}$}}
\put(0,0){\vector(1,1){30}}
\end{picture}
&
\begin{picture}(15,30)(0,0)
\put(10,10){\mbox{$\pi^1_{{\cal M}\pi}$}}
\put(5,30){\vector(0,-1){30}}
\end{picture}
\\
J^1E &
\begin{picture}(30,10)(0,0)
\put(15,6){\mbox{$\widetilde{{\cal F}\Lag}$}}
\put(-2,3){\vector(1,0){40}}
\end{picture}
& {\cal M}\pi
\\
&
\begin{picture}(40,35)(0,0)
 \put(0,35){\vector(1,-2){15}}
 \put(-5,11){\mbox{$\bar\tau^1_0$}}
 \put(38,11){\mbox{$\bar\tau^1$}}
 \put(45,35){\vector(-1,-2){15}}
\end{picture}
&
\\
& M &
\end{array}
\quad
\begin{array}{ccc}
 & &
\begin{picture}(30,10)(0,0)
\put(-60,0)
{\mbox{$(\bar\tau^1\circ\mu)^*\Tan^*M\otimes_{{\cal M}\pi}\Tan{\cal M}\pi$}}
\end{picture}
\\
 & 
\begin{picture}(30,30)(0,0)
\put(-20,19){\mbox{$\tilde\nabla$}}
\put(-26,0){\vector(1,1){30}}
\end{picture}
&
\begin{picture}(15,30)(0,0)
\put(10,10){\mbox{$\kappa_{{\cal M}\pi}$}}
\put(5,30){\vector(0,-1){30}}
\end{picture}
\\
 J^1E &
\begin{picture}(30,10)(0,0)
\put(10,6){\mbox{$\widetilde{{\cal F}\Lag}$}}
\put(-23,3){\vector(1,0){80}}
\end{picture}
& {\cal M}\pi
\\
&
\begin{picture}(80,35)(0,0)
 \put(0,35){\vector(1,-1){35}}
 \put(-5,11){\mbox{$\bar\tau^1_0$}}
 \put(73,11){\mbox{$\bar\tau^1$}}
 \put(85,35){\vector(-1,-1){35}}
\end{picture}
&
\\
& M &
\end{array}
$$

In the same way as classes of locally decomposable and transverse
 $m$-vector fields in a fiber bundle are associated with
 orientable jet fields and connections \cite{EMR-98},
 we have an analogous result in the current situation. In fact:

\begin{teor}
Classes of locally decomposable and $(\bar\tau^1\circ\mu)$-transverse
 $m$-vector fields along $\widetilde{{\cal F}\Lag}$
 are in one-to-one correspondence with
 jet fields along $\widetilde{{\cal F}\Lag}$, and hence with
 Ehresmann connection forms along $\widetilde{{\cal F}\Lag}$.
\label{mvfcon}
\end{teor}
\proof
Bearing in mind (\ref{diagpro1}), in the current situation we have the diagram
\bea
\begin{array}{ccccc}
& \pi^*\Tan^*M\otimes_E\Tan E
&
\begin{picture}(135,20)(0,0)
\put(35,9){\mbox{$\hat\tau\otimes\Tan(\tau^1\circ\mu)$}}
\put(135,3){\vector(-1,0){135}}
\end{picture}
& (\bar\tau^1\circ\mu)^*\Tan^*M\otimes_{{\cal M}\pi}\Tan{\cal M}\pi &
\\
& \begin{picture}(15,35)(0,0)
\put(10,0){\vector(0,1){35}}
\put(-10,12){\mbox{$\Upsilon'_E$}}
\end{picture}
& &
\begin{picture}(15,35)(0,0)
\put(10,0){\vector(0,1){35}}
\put(14,12){\mbox{$\Upsilon'_{{\cal M}\pi}$}}
\end{picture} &
\\
\begin{picture}(5,20)(0,0)
\put(-15,0){\mbox{$\{\Lambda^m\Tan J^1E\}$}}
\end{picture}
 & \supset D^m\Tan E
&
\begin{picture}(135,20)(0,0)
\put(35,9){\mbox{$\{\Lambda^m\Tan(\tau^1\circ\mu)\}$}}
\put(135,3){\vector(-1,0){135}}
\end{picture}
 & D^m\Tan{\cal M}\pi \subset &
\begin{picture}(5,20)(0,0)
\put(-35,0){\mbox{$\{\Lambda^m\Tan{\cal M}\pi\}$}}
\end{picture}
 \\
&
\begin{picture}(15,70)(0,0)
\put(-15,31){\mbox{$\Upsilon_E$}}
\put(3,0){\vector(0,1){70}}
\put(10,31){\mbox{$\Upsilon_E^{-1}$}}
\put(7,70){\vector(0,-1){70}}
\end{picture}
 & 
\begin{picture}(135,70)(0,0)
\put(43,23){\mbox{$j^1(\tau^1\circ\mu)$}}
\put(135,31){\vector(-4,-1){135}}
\end{picture}
&
\begin{picture}(15,70)(0,0)
\put(-19,53){\mbox{$\Upsilon_{{\cal M}\pi}$}}
\put(8,46){\vector(0,1){25}}
\put(14,53){\mbox{$\Upsilon_{{\cal M}\pi}^{-1}$}}
\put(12,71){\vector(0,-1){25}}
\put(-16,8){\mbox{$\pi^1_{{\cal M}\pi}$}}
\put(10,25){\vector(0,-1){25}}
\put(-10,33){\mbox{$J^1{\cal M}\pi$}}
\put(32,10){\mbox{$\{\sigma_{{\cal M}\pi}\}$}}
\put(57,70){\vector(-1,-2){35}}
\end{picture}
&
\\
& J^1E &
\begin{picture}(135,10)(0,0)
\put(63,-15){\mbox{$\widetilde{{\cal F}\Lag}$}}
\put(0,3){\vector(1,0){135}}
\end{picture}
& {\cal M}\pi &
\\
\begin{picture}(10,70)(0,0)
\put(10,0){\vector(0,1){155}}
\put(-5,70){\mbox{$\rho_E$}}
\end{picture}
 & &
\begin{picture}(135,70)(0,0)
\put(7,43){\mbox{$\pi^1$}}
\put(112,43){\mbox{$\tau^1\circ\mu$}}
\put(63,28){\mbox{$E$}}
\put(0,70){\vector(3,-2){55}}
\put(135,70){\vector(-3,-2){55}}
\end{picture}
 &
\begin{picture}(10,70)(0,0)
\put(5,0){\vector(0,1){70}}
\put(-21,27){\mbox{$\sigma_{{\cal M}\pi}$}}
\put(15,0){\vector(1,2){75}}
\put(60,70){\mbox{$\rho_{{\cal M}\pi}$}}
\end{picture}
 &
\\
\Lambda^m\Tan E & &
\begin{picture}(135,10)(0,0)
\put(10,9){\mbox{$\Lambda^m\Tan(\tau^1\circ\mu)$}}
\put(135,3){\vector(-1,0){210}}
\end{picture}
& \Lambda^m\Tan{\cal M}\pi &
\end{array}
\label{above}
\eea
(where the natural projection $\hat\tau\otimes\Tan(\tau^1\circ\mu)$
is defined in a similar way to $\tau\otimes\Tan\pi^1$
in (\ref{mvfcon0})). 

Now, if $\{\tilde{\cal X}\}\colon
J^1E\to D^m\Tan{\cal M}\pi\subset\{\Lambda^m\Tan{\cal M}\pi\}$
is a class of non-vanishing, locally decomposable
and $(\bar\tau^1\circ\mu)$-transverse
$m$-vector fields along $\widetilde{{\cal F}\Lag}$, then
from this class $\{\tilde{\cal X}\}$ we obtain
$\widetilde{{\cal Y}_{\cal X}}=
\Upsilon_{{\cal M}\pi}^{-1}\circ \{\tilde{\cal X}\}$,
and conversely, from a jet field $\tilde{\cal Y}$
along $\widetilde{{\cal F}\Lag}$ we construct
$\{\tilde{\cal X}\}=\Upsilon_{{\cal M}\pi}\circ\tilde{\cal Y}$.
In the same way, from the class $\{\tilde{\cal X}\}$ we obtain
$\widetilde{\nabla_{\cal X}}=\Upsilon_{{\cal M}\pi}'\circ\{\tilde{\cal X}\}$,
and conversely, from an Ehresmann connection form $\tilde\nabla$
along $\widetilde{{\cal F}\Lag}$ we construct
$\{\tilde{\cal X}\}={\Upsilon_{{\cal M}\pi}'}^{-1}\circ\tilde\nabla$.
\qed

The local expression of a representative
of the class $\{\tilde{\cal X}\}$ of non-vanishing, locally decomposable and
$(\bar\tau^1\circ\mu)$-transverse $m$-vector fields along
$\widetilde{{\cal F}\Lag}$, and its
associated jet field $\widetilde{{\cal Y}_{\cal X}}$
 and connection form $\widetilde{\nabla_{\cal X}}$
along $\widetilde{{\cal F}\Lag}$ are
\bea
 \tilde{\cal X} &=&
 \bigwedge_{\alpha=1}^m
 F(x^\nu,y^B,v^B_\nu)
 \left[\left(\derpar{}{x^\alpha}\circ\widetilde{{\cal F}\Lag}\right)+
 f_\alpha^A(x^\nu,y^B,v^B_\nu)
 \left(\derpar{}{y^A}\circ\widetilde{{\cal F}\Lag}\right)+
 \right.
\nonumber
 \\ & & \left.
 g^\eta_{A\alpha}(x^\nu,y^B,v^B_\nu)
\left(\derpar{}{p^\eta_A}\circ\widetilde{{\cal F}\Lag}\right)+
 h_\alpha(x^\nu,y^B,v^B_\nu)
\left(\derpar{}{p}\circ\widetilde{{\cal F}\Lag}\right)\right]
\nonumber
\\
 \widetilde{{\cal Y}_{\cal X}}&=&
\left(x^\alpha,y^A,\derpar{\lag}{v^A_\alpha},
\lag-v^A_\alpha\derpar{\lag}{v^A_\alpha};
f^A_\alpha,g^\eta_{A\alpha},h_\alpha\right)
\label{lekappa}
 \\
\widetilde{\nabla_{\cal X}} &=&
(\d x^\alpha\circ\widetilde{{\cal F}\Lag})\otimes
\left[\left(\derpar{}{x^\alpha}\circ\widetilde{{\cal F}\Lag}\right)+
 f_\alpha^A\left(\derpar{}{y^A}\circ\widetilde{{\cal F}\Lag}\right)+
 g^\eta_{A\alpha}\left(\derpar{}{p^\eta_A}\circ\widetilde{{\cal F}\Lag}\right)+
 h_\alpha\left(\derpar{}{p}\circ\widetilde{{\cal F}\Lag}\right)\right]
\nonumber
\eea

Now, let 
$\{\tilde{\cal X}\}$ be a class of non-vanishing, locally decomposable
and $(\bar\tau^1\circ\mu)$-transverse
$m$-vector fields along $\widetilde{{\cal F}\Lag}$,
let $\widetilde{{\cal Y}_{\cal X}}\colon J^1E\to J^1{\cal M}\pi$ be
its associated jet field along $\widetilde{{\cal F}\Lag}$,
and $\widetilde{\nabla_{\cal X}}\colon J^1E\to
(\bar\tau^1\circ\mu)^*\Tan^*M\otimes_{{\cal M}\pi}\Tan{\cal M}\pi$
its associated Ehresmann connection form along $\widetilde{{\cal F}\Lag}$.
Then bearing in mind the commutativity of
the diagram (\ref{above}) (see also (\ref{abovebis}) below),
and the relations among these elements, we have
$$
j^1(\tau^1\circ\mu)\circ\widetilde{{\cal Y}_{\cal X}}=
\Upsilon_E^{-1}\circ\{\Lambda^m\Tan (\tau^1\circ\mu)\}
\circ\{\tilde{{\cal X}}\}=
\Upsilon_E^{-1}\circ{\Upsilon_E'}^{-1}\circ
\hat\tau\otimes\Tan(\tau^1\circ\mu)\circ\nabla
$$
If $\tilde{\cal X}\colon J^1E\to\Lambda^m\Tan{\cal M}\pi$ is
a representative of the class $\{\tilde{\cal X}\}$, introducing
 \beq
\tilde\varrho_E:=\Upsilon_E^{-1}\circ
\{\Lambda^m\Tan(\tau^1\circ\mu)\}\circ\rho_{{\cal M}\pi}=
\Upsilon_E^{-1}\circ\rho_E\circ\Lambda^m\Tan(\tau^1\circ\mu)
\label{tilderho}
\eeq
we have that
$$
\Upsilon_E^{-1}\circ\{\Lambda^m\Tan (\tau^1\circ\mu)\}
\circ\{\tilde{\cal X}\}=
\Upsilon_E^{-1}\circ\{\Lambda^m\Tan (\tau^1\circ\mu)\}
\circ\rho_{{\cal M}\pi}\circ\tilde{\cal X}\equiv
\tilde\varrho_E\circ\tilde{\cal X}
$$
 from the above equality we obtain
$$
j^1(\tau^1\circ\mu)\circ\widetilde{{\cal Y}_{\cal X}}=
\tilde\varrho_E\circ\tilde{\cal X}=
\Upsilon_E^{-1}\circ{\Upsilon_E'}^{-1}\circ
\hat\tau\otimes\Tan(\tau^1\circ\mu)\circ\nabla
$$

\begin{definition}
$\widetilde{{\cal Y}_{\cal X}}$, its equivalent $\widetilde{\nabla_{\cal X}}$,
 and their associated class
$\{\tilde{\cal X}\}$ are said to be {\rm semi-holonomic} if
$$
\Upsilon_E^{-1}\circ{\Upsilon_E'}^{-1}\circ
\hat\tau\otimes\Tan(\tau^1\circ\mu)\circ\nabla=
\Upsilon_E^{-1}\circ\{\Lambda^m\Tan (\tau^1\circ\mu)\}
\circ\{\tilde{{\cal X}}\}=
j^1(\tau^1\circ\mu)\circ\widetilde{{\cal Y}_{\cal X}}={\rm Id}_{J^1E}
$$
If $\tilde{\cal X}\in\{\tilde{\cal X}\}$ is a representative of this class,
then the above condition leads to
$$
\tilde\varrho_E\circ\tilde{\cal X}=
{\rm Id}_{J^1E}
$$
and $\tilde{\cal X}$ is a {\rm semi-holonomic} $m$-vector field along
$\widetilde{{\cal F}\Lag}$.
\end{definition}

In this case, we have completed the diagram (\ref{above}) as follows
\bea
\begin{array}{ccccc}
& \pi^*\Tan^*M\otimes_E\Tan E
&
\begin{picture}(135,20)(0,0)
\put(35,9){\mbox{$\hat\tau\otimes\Tan(\tau^1\circ\mu)$}}
\put(135,3){\vector(-1,0){135}}
\end{picture}
& (\bar\tau^1\circ\mu)^*\Tan^*M\otimes_{{\cal M}\pi}\Tan{\cal M}\pi &
\\
& \begin{picture}(15,35)(0,0)
\put(10,0){\vector(0,1){35}}
\put(-10,12){\mbox{$\Upsilon'_E$}}
\end{picture}
& &
\begin{picture}(15,35)(0,0)
\put(10,0){\vector(0,1){35}}
\put(14,12){\mbox{$\Upsilon'_{{\cal M}\pi}$}}
\end{picture} &
\\
\begin{picture}(5,20)(0,0)
\put(-15,0){\mbox{$\{\Lambda^m\Tan J^1E\}$}}
\end{picture}
 & \supset D^m\Tan E
&
\begin{picture}(135,20)(0,0)
\put(0,11){\mbox{$\{\Lambda^m\Tan(\tau^1\circ\mu)\}$}}
\put(135,3){\vector(-1,0){135}}
\end{picture}
&
D^m\Tan{\cal M}\pi \subset &
\begin{picture}(5,20)(0,0)
\put(-35,0){\mbox{$\{\Lambda^m\Tan{\cal M}\pi\}$}}
\end{picture}
\\
&
\begin{picture}(15,70)(0,0)
\put(-13,29){\mbox{$\Upsilon_E$}}
\put(5,0){\vector(0,1){70}}
\put(13,29){\mbox{$\Upsilon_E^{-1}$}}
\put(10,70){\vector(0,-1){70}}
\end{picture}
 &
\begin{picture}(135,70)(0,0)
\put(80,117){\mbox{$\widetilde{\nabla_{\cal X}}$}}
\put(0,10){\vector(1,1){125}}
\put(72,55){\mbox{$\{\tilde{\cal X}\}$}}
\put(0,5){\vector(2,1){135}}
\put(75,33){\mbox{$j^1(\tau^1\circ\mu)$}}
\put(135,33){\vector(-4,-1){135}}
\put(90,6){\mbox{$\widetilde{{\cal Y}_{\cal X}}$}}
\put(0,-5){\vector(4,1){135}}
\end{picture}
&
\begin{picture}(15,70)(0,0)
\put(-16,53){\mbox{$\Upsilon_{{\cal M}\pi}$}}
\put(10,46){\vector(0,1){25}}
\put(-16,8){\mbox{$\pi^1_{{\cal M}\pi}$}}
\put(10,25){\vector(0,-1){25}}
\put(-10,33){\mbox{$J^1{\cal M}\pi$}}
\put(32,10){\mbox{$\{\sigma_{{\cal M}\pi}\}$}}
\put(57,70){\vector(-1,-2){35}}
\end{picture}
&
\\
& J^1E &
\begin{picture}(135,10)(0,0)
\put(63,-15){\mbox{$\widetilde{{\cal F}\Lag}$}}
\put(0,3){\vector(1,0){135}}
\end{picture}
& {\cal M}\pi &
\\
\begin{picture}(10,70)(0,0)
\put(10,0){\vector(0,1){155}}
\put(-5,70){\mbox{$\rho_E$}}
\end{picture}
 & &
\begin{picture}(135,70)(0,0)
\put(80,30){\mbox{$\tilde{\cal X}$}}
\put(0,65){\vector(2,-1){135}}
\end{picture}
 &
\begin{picture}(10,70)(0,0)
\put(5,0){\vector(0,1){70}}
\put(-21,27){\mbox{$\sigma_{{\cal M}\pi}$}}
\put(17,0){\vector(1,2){75}}
\put(60,70){\mbox{$\rho_{{\cal M}\pi}$}}
\end{picture}
 &
\\
\Lambda^m\Tan E & &
\begin{picture}(135,10)(0,0)
\put(10,9){\mbox{$\Lambda^m\Tan(\tau^1\circ\mu)$}}
\put(135,3){\vector(-1,0){210}}
\end{picture}
& \Lambda^m\Tan{\cal M}\pi &
\end{array}
\label{abovebis}
\eea

In order to obtain the corresponding local expressions,
consider natural charts of adapted coordinates
$(x^\alpha,y^A,v_\alpha^A)$ and $(x^\alpha,y^A,p^\alpha_A,p)$,
 in $J^1E$ and ${\cal M}\pi$ respectively, and the induced chart
$(x^\alpha,y^A,p_A^\alpha,p;f^A_\alpha ,g^\eta_{A\alpha},h_\alpha)$
in $J^1{\cal M}\pi$. Let
$\tilde{\bf y}\in J^1{\cal M}\pi$ with
\dst\tilde{\bf y}\stackrel{\tau^1_1}{\mapsto}{\bf y}\stackrel{\bar\tau^1}
{\mapsto}x\) .
If $\tilde\psi\colon M\to{\cal M}\pi$ is a representative of ${\bf y}$,
then $\tilde\psi(x)={\bf y}$, and
\beann
{\bf y}&\equiv&(x^\alpha,y^A,p^\alpha_A,p)=
(x^\alpha,\psi^A(x),\psi^\alpha_A(x),\psi(x))\equiv\tilde\psi(x)
\\
\tilde{\bf y}&\equiv&
 (x^\alpha,y^A,p_A^\alpha ,f_\nu^A,g_{A\alpha}^\eta,h_\alpha )=
\left( x^\alpha,\psi^A(x),\psi_A^\alpha(x),
\derpar{{\psi}^A}{x^\alpha}(x),
\derpar{{\psi}_A^\eta}{x^\alpha}(x),\derpar{\psi}{x^\alpha}(x)\right)\equiv
(j^1\tilde\psi)(x)
\eeann
Now we can construct the section $\tau^1\circ\tilde\psi\colon M\to E$,
which is a representative of the point $\bar y\in J^1E$, such that
\dst\bar y\stackrel{\pi^1}{\mapsto}y\stackrel{\pi}{\mapsto}x\) . Thus
$$
\bar y\equiv
(x^\alpha,y^A,v_\alpha^A)=
\left( x^\alpha,\psi^A(x),\derpar{\psi^A}{x^\alpha}(x)\right)
\equiv [j^1(\tau^1\circ\tilde\psi)](x)
$$
On the other hand, recall that the map $j^1\tau^1$ is defined by
$j^1\tau^1(\tilde{\bf y}):=[j^1(\tau^1\circ\tilde\psi )](x)$,
for every section $\tilde\psi$.
Therefore, since $j^1\tau^1(\tilde{\bf y})=\bar y$, we conclude that
$$
j^1\tau^1(x^\alpha,y^A,p^\alpha_A,p;f_\alpha^A,g_{A\alpha}^\eta,h_\alpha)
=(x^\alpha,y^A,f^A_\alpha)
$$
As a consequence, if $\tilde{\cal Y}$ is a jet field along
$\widetilde{{\cal F}\Lag}$, the condition of being semi-holonomic
is locally equivalent to demanding that
\beann
(x^\alpha,y^A,v_\alpha^A) &\equiv& \bar y =
(j^1\tau^1\circ\tilde{\cal Y})(\bar y)\equiv
(j^1\tau^1\circ\tilde{\cal Y})(x^\alpha,y^A,v_\alpha^A)=
(j^1\tau^1\circ\tilde{\cal Y})
\left( x_\alpha,\psi^A(x),\derpar{\psi^A}{x^\alpha}(x)\right)
\\ &=&
j^1\tau^1\left( x^\alpha,{\psi}^A(x),{\psi}_A^\alpha(x),
\derpar{{\psi}^A}{x^\alpha}(x),
\derpar{{\psi}_A^\eta}{x^\alpha}(x),\derpar{\psi}{x^\alpha}(x)\right)
\\ &=&
\left( x^\alpha,{\psi}^A(x),\derpar{{\psi}^A}{x^\alpha}(x)\right)=
(x^\alpha,y^A,f^A_\alpha)
\eeann
that is, $f^A_\alpha=v_\alpha^A$ (in (\ref{lekappa})).

Now, the generalization of the integrability conditions
(\ref{int00}) and (\ref{int0})
 to the current situation leads to the following:

\begin{definition}
Let $\tilde{\cal X}$ be a non-vanishing and locally decomposable
$m$-vector field along $\widetilde{{\cal F}\Lag}$.
 A section $\varphi\colon M\to J^1E$ is said to be an {\rm integral section}
 of $\tilde{\cal X}$ if
\beq
\Lambda^m\Tan (\widetilde{{\cal F}\Lag}\circ\varphi)=
f\tilde{\cal X}\circ\varphi\circ\sigma_M
\label{int1}
\eeq
where $f\in\Cinfty (J^1E)$ is a non-vanishing function.
 $\tilde{\cal X}$ is said to be
 {\rm integrable} if it admits integral sections.
Thus, we have the diagram
$$
\begin{array}{ccccc}
\Lambda^m\Tan M &
\begin{picture}(55,20)(0,0)
\put(27,8){\mbox{$\Lambda^m\Tan (\widetilde{{\cal F}\Lag}\circ\varphi)$}}
\put(0,3){\vector(1,0){150}}
\end{picture} & & &
\Lambda^m\Tan{\cal M}\pi
\\
\begin{picture}(10,35)(0,0)
\put(-15,15){\mbox{$\sigma_M$}}
\put(5,35){\vector(0,-1){35}}
\end{picture}
& & &
\begin{picture}(55,35)(0,0)
\put(7,15){\mbox{$f\tilde{\cal X}$}}
\put(0,0){\vector(3,2){55}}
\end{picture}
 &
\begin{picture}(15,35)(0,0)
\put(13,15){\mbox{$\sigma_{{\cal M}\pi}$}}
\put(10,35){\vector(0,-1){35}}
\end{picture}
\\
M &
\begin{picture}(55,10)(0,0)
\put(21,6){\mbox{$\varphi$}}
\put(0,3){\vector(1,0){55}}
\end{picture}
& J^1E &
\begin{picture}(55,10)(0,0)
\put(20,6){\mbox{$\widetilde{{\cal F}\Lag}$}}
\put(0,3){\vector(1,0){60}}
\end{picture}
& {\cal M}\pi
\end{array}
$$
 $\tilde{\cal X}$ is said to be
 {\rm holonomic} if its integral sections
 are holonomic, that is, $\varphi =j^1\phi$ for some section
 $\phi\colon M\to E$.
\label{int11}
\end{definition}

Observe that we are characterizing the
integrability of the entire class $\{\tilde{\cal X}\}$.
Note also that this definition of integral section
is equivalent to stating that the image of the section
$\tilde\varphi=
\widetilde{{\cal F}\Lag}\circ\varphi\colon M\to\Lambda^m\Tan{\cal M}\pi$
is an integral submanifold of the distribution
${\cal D}(\tilde{\cal X})$; that is,
$\Tan_{\tilde\varphi(x)}({\rm Im}\,\tilde\varphi)=
[{\cal D}(\tilde{\cal X})]_{\varphi(x)}$,
for every $x\in M$.
Finally, it is important to remark that,
if a $m$-vector field along $\widetilde{{\cal F}\Lag}$
is not integrable everywhere in $J^1E$, it could be integrable
on a submanifold ${\cal I}\hookrightarrow J^1E$
(see the comment at the end of Section \ref{lf}).

\begin{quote}
{\bf Remark}:
Of course, a class $\{\tilde{\cal X}\}$
of $m$-vector fields along $\widetilde{{\cal F}\Lag}$
is integrable (resp. holonomic) if, and only if,
its associated jet field $\widetilde{{\cal Y}_{\cal X}}$ and connection
$\tilde\nabla$ along $\widetilde{{\cal F}\Lag}$ are also.

In addition, the class $\{\tilde{\cal X}\}$ and
its associated jet field $\tilde{\cal Y}$ and connection
$\tilde\nabla$ are holonomic if, and only if,
they are integrable and semi-holonomic
(see the proof for multivector fields, jet fields and connections
 in jet bundles in \cite{EMR-98}).
\end{quote}

In a system of natural coordinates, if
 $\varphi(x)=(x^\alpha ,\varphi^A(x),\varphi_\alpha^A(x))$
 is an integral section of $\tilde{\cal X}$
(see (\ref{lekappa}) for its local expression), then
the following system of partial differential equations holds
\bea
f^A_\alpha(x^\beta,\varphi^C(x),\varphi^C_\beta(x)) &=&
\derpar{\varphi^A}{x^\alpha}
\nonumber \\
g^\eta_{A_\alpha}(x^\beta,\varphi^C(x),\varphi^C_\beta(x)) &=&
\frac{\partial^2\lag}{\partial x^\alpha\partial v^A_\eta}+
\frac{\partial^2\lag}{\partial y^B\partial v^A_\eta}
\derpar{\varphi^B}{x^\alpha}+
\frac{\partial^2\lag}{\partial v^B_\nu\partial v^A_\eta}
\derpar{\varphi^B_\nu}{x^\alpha}
 \label{intseck}
 \\
h_\alpha(x^\beta,\varphi^C(x),\varphi^C_\beta(x)) &=&
\derpar{\lag}{x^\alpha}+\derpar{\lag}{y^A}\derpar{\varphi^A}{x^\alpha}-
\varphi^A_\eta\left(
\frac{\partial^2\lag}{\partial x^\alpha\partial v^A_\eta}+
\frac{\partial^2\lag}{\partial y^B\partial v^A_\eta}
\derpar{\varphi^B}{x^\alpha}+
\frac{\partial^2\lag}{\partial v^B_\nu\partial v^A_\eta}
\derpar{\varphi^B_\nu}{x^\alpha}\right)
\nonumber
\eea
In particular, if $\varphi$ is holonomic and $\varphi=j^1\phi$
with $\phi(x)=(x^\alpha ,\phi^A(x))$, then $\varphi^A\equiv\phi^A$, and
\dst\varphi_\alpha^A=\derpar{\phi^A}{x^\alpha}\)~. Therefore
the above system is second order.

As a final remark, we define the
contraction of jet fields along $\widetilde{{\cal F}\Lag}$ with
differential forms along $\widetilde{{\cal F}\Lag}$, as
a natural extension of the same operation between
jet fields and differential forms in a jet bundle. Thus,
let $\tilde{\cal Y}$ be a jet field along $\widetilde{{\cal F}\Lag}$, then
to every $Z\in\vf (M)$ we can associate a vector field $\tilde{\cal Z}$ along
$\widetilde{{\cal F}\Lag}$, which is given by
$$
\tilde{\cal Z}(\bar y):=(\Tan_x\psi )(Z_x)
$$
for every $\bar y\in J^1E$, with
 $\bar y\stackrel{\pi^1}{\to}y\stackrel{\pi}{\mapsto}x$,
 and $\psi\in\tilde{\cal Y}(\bar y)$.
If \dst Z=F^\alpha\derpar{}{x^\alpha}\)
the local expression of $\tilde{\cal Z}$ is
$$
\tilde{\cal Z}=
F^\alpha\left[\left(\derpar{}{x^\alpha}\circ\widetilde{{\cal F}\Lag}\right)+
 f_\alpha^A\left(\derpar{}{y^A}\circ\widetilde{{\cal F}\Lag}\right)+
 g^\eta_{A\alpha}\left(\derpar{}{p^\eta_A}\circ\widetilde{{\cal F}\Lag}\right)+
 h_\alpha\left(\derpar{}{p}\circ\widetilde{{\cal F}\Lag}\right)\right]
$$

\begin{definition}
Let ${\mit \Xi}$ be a $(m+p)$-form along $\widetilde{{\cal F}\Lag}$
(with $p\geq 0$), then $\inn (\tilde{\cal Y}){\mit \Xi}$ is an element of
$\df^m(M,\bar\tau^1\circ\mu)\otimes_{{\cal M}\pi}
\df^p({\cal M}\pi,\widetilde{{\cal F}\Lag})$ defined as
$$
[(\inn (\tilde{\cal Y}){\mit \Xi})
(\bar y;Z_1,\ldots ,Z_m;\tilde {\cal X}_1,\ldots ,\tilde {\cal X}_p):=
{\mit \Xi} (\bar y;\tilde{\cal Z}_1,\ldots ,
\tilde{\cal Z}_m,\tilde {\cal X}_1,\ldots ,\tilde {\cal X}_p)
$$
for $Z_1,\ldots ,Z_m\in\vf (M)$, and
$\tilde {\cal X}_1,\ldots ,\tilde {\cal X}_p$
vector fields in ${\cal M}\pi$ along $\widetilde{{\cal F}\Lag}$.

This map is extended by zero to forms of degree $k<m$, and
it is a $\Cinfty (M)$-linear and alternate on
$Z_1,\ldots ,Z_m$ and $\tilde {\cal X}_1,\ldots ,\tilde {\cal X}_p$.
\label{innerjet}
\end{definition}

\begin{quote}
{\bf Remark}:
 Observe that contracting a form ${\mit \Xi}$ along $\widetilde{{\cal F}\Lag}$
with a jet field $\tilde{\cal Y}$ along $\widetilde{{\cal F}\Lag}$
is equivalent to contracting ${\mit \Xi}$ with a (suitable) representative of
the class $\{\tilde{\cal X}\}$ of $m$-vector fields
along $\widetilde{{\cal F}\Lag}$ associated with $\tilde{\cal Y}$.
\end{quote}

 As is evident, all the definitions and results in this section
 can also be stated in a similar way
 for $m$-vector fields ${\cal X}$ along the restricted Legendre map 
 ${\cal F}\Lag$, and orientable jet fields ${\cal Y}$
 and connection forms $\nabla$ along ${\cal F}\Lag$.

\subsection{The extended field operators}

Let $\ls$ be a Lagrangian system. Then:

\begin{definition}
\ben
\item
An {\rm extended $m$-vector field operator} $\tilde{\cal K}$ associated
with $\ls$ is a map
$\tilde{\cal K}\colon J^1E \longrightarrow \Lambda^m\Tan{\cal M}\pi$
verifying the following conditions:
\ben
\item
 ({\sl Structural condition}):
 $\tilde{\cal K}$ is a non-vanishing, locally decomposable
 and $(\bar\tau^1\circ\mu)$-transverse
 $m$-vector field along $\widetilde{{\cal F}\Lag}$.
\item
({\sl Field equation condition}):
$\widetilde{{\cal F}\Lag}^*
[\inn(\tilde{\cal K})(\Omega\circ\widetilde{{\cal F}\Lag})]=0$.
\item
({\sl Semi-holonomy condition}):
 $\tilde{\cal K}$ is semi-holonomic.
\een
\item
An {\rm extended jet field operator} $\widetilde{{\cal Y}_{\cal K}}$ associated
with $\ls$ is a map
$\widetilde{{\cal Y}_{\cal K}}\colon J^1E \longrightarrow J^1{\cal M}\pi$
verifying the following conditions:
\ben
\item
 ({\sl Structural condition}):
 $\widetilde{{\cal Y}_{\cal K}}$ is an orientable jet field along
 $\widetilde{{\cal F}\Lag}$.
\item
({\sl Field equation condition}):
$\widetilde{{\cal F}\Lag}^*
[\inn(\widetilde{{\cal Y}_{\cal K}})(\Omega\circ\widetilde{{\cal F}\Lag})]=0$.
\item
({\sl Semi-holonomy condition}):
 $\widetilde{{\cal Y}_{\cal K}}$ is semi-holonomic.
\een
\item
An {\rm extended connection operator} $\widetilde{\nabla_{\cal K}}$ associated
with $\ls$ is a map
$\widetilde{\nabla_{\cal K}}\colon J^1E \longrightarrow
(\bar\tau^1\circ\mu)^*\Tan^*M\otimes_{{\cal M}\pi}\Tan{\cal M}\pi$
verifying the following conditions:
\ben
\item
 ({\sl Structural condition}):
 $\widetilde{\nabla_{\cal K}}$ is an orientable Ehresmann connection form along
 $\widetilde{{\cal F}\Lag}$.
\item
({\sl Field equation condition}):
$\widetilde{{\cal F}\Lag}^*
[\inn(\widetilde{\nabla_{\cal K}})(\Omega\circ\widetilde{{\cal F}\Lag})-
(m-1)(\Omega\circ\widetilde{{\cal F}\Lag})]=0$.
\item
({\sl Semi-holonomy condition}):
 $\widetilde{\nabla_{\cal K}}$ is semi-holonomic.
\een
\een
\label{tildek}
\end{definition}

\begin{quote}
{\bf Remark}:
Note that the field equation condition of the first item in
this definition defines
not a single $m$-vector field along $\widetilde{{\cal F}\Lag}$,
but classes of them. A representative can be selected by adding
the following {\sl normalization condition}
\beq
\inn(\tilde{\cal K})[(\bar\tau^1\circ\mu)^*\omega\circ
\widetilde{{\cal F}\Lag})]=1
\label{normcon}
\eeq
which, in its turn, implies the $(\bar\tau^1\circ\mu)$-transversality
condition.
\end{quote}

\begin{teor}
A class of extended $m$-vector field operators $\tilde{\cal K}$ is associated
 with an extended jet field operator $\widetilde{{\cal Y}_{\cal K}}$
and an extended connection operator $\widetilde{\nabla_{\cal K}}$,
 and conversely.
\label{efocon}
\end{teor}
\proof
It follows from Theorem \ref{mvfcon}.
The only point to be proved is the equivalence between the
field equation conditions, which follows after a simple calculation in
coordinates, using the local expressions (\ref{lekappa}).
\qed

\begin{teor}
{\rm (Existence and local multiplicity}).
There exist classes of extended $m$-vector field operators $\tilde{\cal K}$
 for $\ls$, and hence there exist also
 extended jet field operators $\widetilde{{\cal Y}_{\cal K}}$
and extended connection operators $\widetilde{\nabla_{\cal K}}$.
In a local system, they depend on $N(m^2-1)$ arbitrary functions.
\label{holsecreg}
\end{teor}
\proof
First we analyze the local existence, and then their global extension.

Let $\bar y\equiv (x^\alpha,y^A,v^A_\alpha)\in J^1E$, and
$\hat{\bf y}=({\bf y},X_{{\bf y}})\in\Lambda^m\Tan{\cal M}\pi$
(where ${\bf y}\equiv (x^\alpha,y^A,p_A^\alpha,p)\in\Tan{\cal M}\pi$
and $X_{{\bf y}}\in\Lambda^m\Tan_{{\bf y}}{\cal M}\pi$),
 such that $\tilde{\cal K}(\bar y)=\hat{\bf y}$.
 First, the equation
 $\sigma_{{\cal M}\pi}\circ\tilde{\cal K}=\widetilde{{\cal F}\Lag}$
 in condition $1$ implies that
 ${\bf y}=\widetilde{{\cal F}\Lag}(\bar y)$, thus
$$
 p_\alpha^A=\derpar{\lag}{v_\alpha^A} \quad , \quad
 p=\lag-v_\alpha^A\derpar{\lag}{v_\alpha^A}
$$
On the other hand, from condition $1$, $X_{{\bf y}}$ is a locally decomposable
and $(\bar\tau^1\circ\mu)$-transverse $m$-vector at ${\bf y}$, hence
$$
 X_{{\bf y}}=\bigwedge_{\alpha=1}^m F(\bar y)
 \left(\derpar{}{x^\alpha}\Big\vert_{\widetilde{{\cal F}\Lag}(\bar y)}+
 f_\alpha^A(\bar y)\derpar{}{y^A}
 \Big\vert_{\widetilde{{\cal F}\Lag}(\bar y)}+
 g^\eta_{A\alpha}(\bar y)\derpar{}{p^\eta_A}
 \Big\vert_{\widetilde{{\cal F}\Lag}(\bar y)}+
 h_\alpha(\bar y)\derpar{}{p}\Big\vert_{\widetilde{{\cal F}\Lag}(\bar y)}
\right)
$$
In this way we can write
\beann
\tilde{\cal K} &=& \bigwedge_{\alpha=1}^m F
 \left[\left(\derpar{}{x^\alpha}\circ\widetilde{{\cal F}\Lag}\right)+
 f_\alpha^A\left(\derpar{}{y^A} \circ\widetilde{{\cal F}\Lag}\right)+
 g^\eta_{A\alpha}\left(\derpar{}{p^\eta_A}
 \circ\widetilde{{\cal F}\Lag}\right)+
 h_\alpha\left(\derpar{}{p}\circ\widetilde{{\cal F}\Lag}\right) \right]
\eeann
Now, the semi-holonomy condition implies that
$f_\alpha^A=v_\alpha^A$.

Next, taking into account (\ref{Omegah}), we have that
\beann
\widetilde{{\cal F}\Lag}^*
[\inn(\tilde{\cal K})(\Omega\circ\widetilde{{\cal F}\Lag})]&=&
(-1)^{m(m+1)/2}F\widetilde{{\cal F}\Lag}^*
[-v^A_\alpha(\d p_A^\alpha\circ\widetilde{{\cal F}\Lag})-
(\d p\circ\widetilde{{\cal F}\Lag})+
g^\alpha_{A\alpha}(\d y^A\circ\widetilde{{\cal F}\Lag})
\\ & &
+\sum_{\eta\not=\alpha}(-1)^\eta
(g^\eta_{A\eta}v^A_\alpha-g^\eta_{A\alpha}v^A_\eta)
(\d x^\alpha\circ\widetilde{{\cal F}\Lag})
+h_\alpha
(\d x^\alpha\circ\widetilde{{\cal F}\Lag})]
\\ &=&
(-1)^{m(m+1)/2}F
\left[-v^A_\alpha\d\left(\derpar{\lag}{v^A_\alpha}\right)-
\d\left(\lag-v^A_\alpha\derpar{\lag}{v^A_\alpha}\right)+
g^\alpha_{A\alpha}(\d y^A\circ\widetilde{{\cal F}\Lag})
\right. \\ & & \left.
+\sum_{\eta\not=\alpha}(-1)^\eta
(g^\eta_{A\eta}v^A_\alpha-g^\eta_{A\alpha}v^A_\eta)
(\d x^\alpha\circ\widetilde{{\cal F}\Lag})
+h_\alpha
(\d x^\alpha\circ\widetilde{{\cal F}\Lag})]
\right]
\\ &=&
(-1)^{m(m-1)/2}F\left[
\left(g^\alpha_{A\alpha}-\derpar{\lag}{y^A} \right)\d y^A+
\right.
 \\ & & \left.
\left(-\derpar{\lag}{x^\alpha}
+\sum_{\eta\not=\alpha}(-1)^\eta
(g^\eta_{A\eta}v^A_\alpha-g^\eta_{A\alpha}v^A_\eta)
+h_\alpha\right)\d x^\alpha\right]
\eeann
and then, from the field equation condition we obtain
\beq
 g^\alpha_{A\alpha}=\derpar{\lag}{y^A}
\quad , \quad
h_\alpha=\derpar{\lag}{x^\alpha}
-\sum_{\eta\not=\alpha}(-1)^\eta
(g^\eta_{A\eta}v^A_\alpha-g^\eta_{A\alpha}v^A_\eta)
\label{coefk}
\eeq
Finally, if we apply the normalization condition (\ref{normcon}), we can choose $F=1$.
In this way we have obtained, for this representative, the local expression
\bea
\tilde{\cal K} &=&
 \bigwedge_{\alpha=1}^m
 \left[\left(\derpar{}{x^\alpha}\circ\widetilde{{\cal F}\Lag}\right)+
 v_\alpha^A\left(\derpar{}{y^A}
 \circ\widetilde{{\cal F}\Lag}\right)+
 g^\eta_{A\alpha}\left(\derpar{}{p^\eta_A}
\circ\widetilde{{\cal F}\Lag}\right)+
\right. \nonumber \\ & & \left.
 \left(\derpar{\lag}{x^\alpha}
-\sum_{\eta\not=\alpha}(-1)^\eta
(g^\eta_{A\eta}v^A_\alpha-g^\eta_{A\alpha}v^A_\eta)\right)
\left(\derpar{}{p}\circ\widetilde{{\cal F}\Lag}\right)
 \right]
\label{tildekcor}
\eea
So, $\tilde{\cal K}$ is determined by the $Nm^2$ coefficients $g^\eta_{A\alpha}$,
which are related  by the first group of
 $N$ independent equations (\ref{coefk}).
Therefore, there are $N(m^2-1)$ arbitrary functions.

These results allow us to assure the local existence of
classes of extended $m$-vector field operators $\tilde{\cal K}$
satisfying the desired conditions.
 The corresponding global solutions
are then obtained using a partition of unity subordinated
to a covering of $J^1E$ made of local natural charts
$\{ U_i\}$. Then, let $\tilde{\cal K}_i$ be the
field operator in the corresponding open set $U_i$.
As every local class $\{\tilde{\cal K}_i\}$
is associated with a local
Ehresmann connection form $\widetilde{\nabla_{\cal K}}_i$
along $\widetilde{{\cal F}\Lag}$ (by Theorem \ref{efocon}),
and the convex combination of connection forms
gives a connection form,
$\widetilde{\nabla_{\cal K}}=g^i\widetilde{\nabla_{\cal K}}_i$ is a (global)
Ehresmann connection form along $\widetilde{{\cal F}\Lag}$
which is associated with the corresponding class.
The class $\{\tilde{\cal K}\}$ associated with
$\widetilde{\nabla_{\cal K}}$ is the global solution.
In an analogous way, taking into account the affine structure of the fibers
of $J^1{\cal M}\pi$, it is meaningful to construct convex combinations
of sections $\widetilde{{\cal Y}_{\cal K}}_i$, so we can define
a global jet field 
$\widetilde{{\cal Y}_{\cal K}}:=g^i\widetilde{{\cal Y}_{\cal K}}_i$
along $\widetilde{{\cal F}\Lag}$ associated with the class
$\{\tilde{\cal K}\}$.

These elements satisfy the conditions of Definition \ref{tildek}.
In particular:
\bit
\item
The classes $\{\tilde{\cal K}\}$ are made of
 non-vanishing, locally decomposable and $(\bar\tau^1\circ\mu)$-transverse
 $m$-vector fields along $\widetilde{{\cal F}\Lag}$,
since they are associated with orientable jet fields and connections
along $\widetilde{{\cal F}\Lag}$.
\item
The field equation condition holds for every $\tilde{\cal K}$,
because it holds for every $\tilde{\cal K}_i$,
and $\tilde{\cal K}$ is a linear combination
 $\tilde{\cal K}=f^i\tilde{\cal K}_i$.
As a consequence, the equivalent field equation conditions
hold for $\widetilde{\nabla_{\cal K}}$ and $\widetilde{{\cal Y}_{\cal K}}$.
\item
The semi-holonomy of $\tilde{\cal K}$ is proved
starting from the semi-holonomy of $\tilde{\cal K}_i$,
and using that $\{{\cal K}\}$ is associated with
$\widetilde{{\cal Y}_{\cal K}}:=g^i\widetilde{{\cal Y}_{\cal K}}_i$,
which is semi-holonomic because so are $\widetilde{{\cal Y}_{\cal K}}_i$.
\qed
\eit

{\bf Remarks}:
\bit
\item
Observe that the existence of these
extended field operators does not depend on
the regularity of the Lagrangian system.
\item
The class $\tilde{\cal K}$
(and hence the associated $\widetilde{{\cal Y}_{\cal K}}$ and
$\widetilde{\nabla_{\cal K}}$) is {\rm integrable} if Definition \ref{int11}
holds for it. Observe that, if $\tilde{\cal K}$,
and hence the associated $\widetilde{{\cal Y}_{\cal K}}$
and $\widetilde{\nabla_{\cal K}}$ are integrable,
they are holonomic, since they are semi-holonomic.
\eit

Among the multiplicity of extended field operators,
we will mainly be interested in those which are integrable,
and hence holonomic. Thus, if 
\dst\varphi\equiv j^1\phi(x)=
\left( x^\mu,\phi^A,\derpar{\phi^A}{x^\nu}\right)\)
is an integral section, then
\dst v^A_\alpha=\derpar{\phi^A}{x^\alpha}\) , and
from (\ref{intseck}) and (\ref{coefk}) we obtain that $\phi$
must be a solution of the following system
$$
\frac{\partial^2\lag}{\partial x^\alpha\partial v^A_\alpha}+
\frac{\partial^2\lag}{\partial y^B\partial v^A_\alpha}
\derpar{\phi^B}{x^\alpha}+
\frac{\partial^2\lag}{\partial v^B_\nu\partial v^A_\alpha}
\frac{\partial^2\phi^B}{\partial x^\alpha\partial x^\nu} =
\derpar{\lag}{y^A}
$$
which are just the Euler-Lagrange equations for $\phi$
(see Theorem \ref{inteor} for a precise statement of
this comment).

In general, we know there is no way of assuring that
this system is integrable in $J^1E$ (even in the hyper-regular case).
In the most favourable cases,
a submanifold ${\cal I}\hookrightarrow J^1E$ could exist
such that there are (classes of) integrable $m$-vector field
operators $\tilde{\cal K}$ on ${\cal I}$ wich are tangent to ${\cal I}$.
As a consequence, the above system is integrable on ${\cal I}$
and the corresponding integral sections solution are in ${\cal I}$.
(See also \cite{EMR-98} and \cite{EMR-99b} for a discussion of
the integrability of multivector fields).

\subsection{The restricted field operators}
\protect\label{rmvfo}

In field theory there is another kind of field operator
which can be defined, and which is justified
because the multimomentum bundle
where the Hamiltonian formalism of field theories takes place is
really $J^{1*}E$, instead of ${\cal M}\pi$,
and hence, in this case, the relevant Legendre map is ${\cal F}\Lag$
instead of $\widetilde{{\cal F}\Lag}$.

\begin{definition}
\ben
\item
Given an extended $m$-vector field operator $\tilde{\cal K}$,
the {\rm restricted $m$-vector field operator} ${\cal K}$
associated with $\tilde{\cal K}$ is
$$
{\cal K}:=\Lambda^m\Tan\mu\circ\tilde{\cal K}
$$
\item
Given an extended jet field operator $\widetilde{{\cal Y}_{\cal K}}$,
the {\rm restricted jet field operator} ${\cal Y}_{\cal K}$
associated with $\widetilde{{\cal Y}_{\cal K}}$ is
$$
{\cal Y}_{\cal K}:=j^1\mu\circ\widetilde{{\cal Y}_{\cal K}}
$$
\item
Given an extended connection operator $\widetilde{\nabla_{\cal K}}$,
the {\rm restricted connection operator} $\nabla_{\cal K}$
associated with $\widetilde{\nabla_{\cal K}}$ is
$$
\nabla_{\cal K}:=(\tau_\mu\otimes\Tan\mu)\circ\widetilde{\nabla_{\cal K}}
$$
\een
\label{restk}
\end{definition}

 So, we have the diagrams
$$
\begin{array}{ccc}
 & & \Lambda^m\Tan{\cal M}\pi
\\
 & 
\begin{picture}(30,60)(0,0)
\put(8,40){\mbox{$\tilde{\cal K}$}}
\put(0,0){\vector(1,2){30}}
\put(15,20){\mbox{${\cal K}$}}
\put(0,-6){\vector(1,1){30}}
\end{picture}
&
\begin{picture}(15,40)(0,0)
\put(10,8){\mbox{$\sigma_{J^{1*}E}$}}
\put(5,20){\vector(0,-1){20}}
\put(10,47){\mbox{$\Lambda^m\Tan\mu$}}
\put(5,60){\vector(0,-1){20}}
\put(-15,28){\mbox{$\Lambda^m\Tan J^{1*}E$}}
\end{picture}
\\
J^1E &
\begin{picture}(30,10)(0,0)
\put(10,6){\mbox{${\cal F}\Lag$}}
\put(0,3){\vector(1,0){40}}
\end{picture}
& J^{1*}E
\end{array}
\quad  \quad
\begin{array}{ccc}
 & & J^1{\cal M}\pi
\\
 & 
\begin{picture}(30,60)(0,0)
\put(0,40){\mbox{$\widetilde{{\cal Y}_{\cal K}}$}}
\put(0,0){\vector(1,2){30}}
\put(13,15){\mbox{${\cal Y}_{\cal K}$}}
\put(8,-6){\vector(1,1){30}}
\end{picture}
&
\begin{picture}(15,40)(0,0)
\put(10,8){\mbox{$\pi^1_{J^{1*}E}$}}
\put(5,20){\vector(0,-1){20}}
\put(10,47){\mbox{$j^1\mu$}}
\put(5,60){\vector(0,-1){20}}
\put(-15,28){\mbox{$J^1J^{1*}E$}}
\end{picture}
\\
J^1E &
\begin{picture}(30,10)(0,0)
\put(15,6){\mbox{${\cal F}\Lag$}}
\put(0,3){\vector(1,0){35}}
\end{picture}
& J^{1*}E
\end{array}
\quad  \quad
\begin{array}{ccc}
 & & (\bar\tau^1\circ\mu)^*\Tan^*M\otimes_{{\cal M}\pi}\Tan{\cal M}\pi
\\
 & 
\begin{picture}(30,60)(0,0)
\put(0,40){\mbox{$\widetilde{\nabla_{\cal K}}$}}
\put(0,0){\vector(1,2){30}}
\put(15,12){\mbox{$\nabla_{\cal K}$}}
\put(0,-6){\vector(2,1){50}}
\end{picture}
&
\begin{picture}(15,40)(0,0)
\put(10,8){\mbox{$\kappa_{J^{1*}E}$}}
\put(5,20){\vector(0,-1){20}}
\put(10,47){\mbox{$\tau_\mu\otimes\Tan\mu$}}
\put(5,60){\vector(0,-1){20}}
\put(-47,28){\mbox{$\bar\tau^{1*}\Tan^*M\otimes_{J^{1*}E}\Tan J^{1*}E$}}
\end{picture}
\\
J^1E &
\begin{picture}(30,10)(0,0)
\put(28,6){\mbox{${\cal F}\Lag$}}
\put(0,3){\vector(1,0){70}}
\end{picture}
& J^{1*}E
\end{array}
$$
(where the natural projection $\tau_\mu\otimes\Tan\mu$
is defined in a similar way to $\tau\otimes\Tan\pi^1$
in (\ref{mvfcon0})). 

The restricted field operators, ${\cal K}$, ${\cal Y}_{\cal K}$, and
 $\nabla_{\cal K}$ are {\sl $m$-vector fields}, {\sl jet fields} and
{\sl Ehresmann connection forms along the  Legendre map} ${\cal F}\Lag$,
respectively. In particular, it is
obvious that every restricted $m$-vector field operator ${\cal K}$ is
non-vanishing, locally decomposable and $\bar\tau^1$-transverse.

\begin{quote}
{\bf Remark}:
In an analogous way to Theorems
\ref{mvfcon} and \ref{efocon}, we can prove that 
for every class of extended $m$-vector field operators $\{\tilde{\cal K}\}$,
 and its associated
$\widetilde{{\cal Y}_{\cal K}}$ and $\widetilde{\nabla_{\cal K}}$,
the class of restricted $m$-vector field operators
$\{{\cal K}\}:=\{\Lambda^m\Tan\mu\}\circ\{\tilde{\cal K}\}$ 
is associated with
 ${\cal Y}_{\cal K}=j^1\mu\circ\widetilde{{\cal Y}_{\cal K}}$,
and $\nabla_{\cal K}=\tau_\mu\otimes\Tan\mu\circ\widetilde{\nabla_{\cal K}}$.
\end{quote}

\begin{prop}
\ben
\item
$\{{\cal K}\}$ and its associated
${\cal Y}_{\cal K}$ and $\nabla_{\cal K}$
are semi-holonomic.
\item
 ${\cal K}$ is integrable if, and only if,
the corresponding $\tilde{\cal K}$ is also.
That is, $\varphi\colon M\to  J^1E$
 is an integral section of $\{{\cal K}\}$ if, and only if,
 it is an integral section of $\{\tilde{\cal K}\}$ too.
\item
${\cal Y}_{\cal K}$ and $\nabla_{\cal K}$
are integrable if, and only if, the corresponding
$\widetilde{{\cal Y}_{\cal K}}$ and $\widetilde{\nabla_{\cal K}}$ are also.
\een
\label{insecrestk}
\end{prop}
\proof
\ben
\item
For every
 $\tilde{\cal K}\in\{\tilde{\cal K}\}$,
and ${\cal K}=\Lambda^m\Tan\mu\circ\tilde{\cal K}$, we have that
$$
\Upsilon_E^{-1}\circ\rho_E\circ\Lambda^m\Tan\tau\circ{\cal K}=
\Upsilon_E^{-1}\circ\rho_E\circ\Lambda^m\Tan\tau
\circ\Lambda^m\Tan\mu\circ\tilde{\cal K}=
\Upsilon_E^{-1}\circ\rho_E\circ\Lambda^m\Tan(\tau\circ\mu)\circ\tilde{\cal K}=
{\rm Id}_{J^1E}
$$
\item
Let $\varphi\colon M\to J^1E$ be a section. We have
$$
\Lambda^m\Tan{\cal F}\Lag\circ\Lambda^m\Tan\varphi=
\Lambda^m\Tan(\mu\circ\widetilde{{\cal F}\Lag})\circ\Lambda^m\Tan\varphi=
\Lambda^m\Tan\mu\circ\Lambda^m\Tan\widetilde{{\cal F}\Lag}
\circ\Lambda^m\Tan\varphi
$$
and, on the other hand, for every $\tilde{\cal K}\in\{\tilde{\cal K}\}$,
and ${\cal K}=\Lambda^m\Tan\mu\circ\tilde{\cal K}$, we have:
$$
{\cal K}\circ\varphi\circ\sigma_M=
\Lambda^m\Tan\mu\circ\tilde{\cal K}\circ\varphi\circ\sigma_M
$$
therefore, if $f\in\Cinfty (J^1E)$ is a non-vanishing function, then
$$
\Lambda^m\Tan\widetilde{{\cal F}\Lag}\circ\Lambda^m\Tan\varphi=
f\tilde{\cal K}\circ\varphi\circ\sigma_M
\ \Longrightarrow \
\Lambda^m\Tan{\cal F}\Lag\circ\Lambda^m\Tan\varphi=
f{\cal K}\circ\varphi\circ\sigma_M
$$
Conversely, if the last relation holds, then the first one is true
for some non-vanishing function $g\in\Cinfty (J^1E)$.
\item
It is a straighforward consequence of all the above results.
\qed
\een

 The coordinate expressions of these elements are
\bea
{\cal K} &=&  \bigwedge_{\alpha=1}^m F
 \left[\left(\derpar{}{x^\alpha}\circ{\cal F}\Lag\right)+
 v_\alpha^A\left(\derpar{}{y^A} \circ{\cal F}\Lag\right)+
 g^\eta_{A\alpha}\left(\derpar{}{p^\eta_A}\circ{\cal F}\Lag\right)\right]
\label{kcor}
\\
{\cal Y}_{\cal K} &=&
\left(x^\alpha,y^A,\derpar{\lag}{v^A_\alpha};v^A_\alpha,g^\eta_{A\alpha}\right)
\nonumber \\
\nabla_{\cal K} &=&  (\d x^\alpha\circ{\cal F}\Lag)\otimes
 \left[\left(\derpar{}{x^\alpha}\circ{\cal F}\Lag\right)+
 v_\alpha^A\left(\derpar{}{y^A} \circ{\cal F}\Lag\right)+
 g^\eta_{A\alpha}\left(\derpar{}{p^\eta_A}\circ{\cal F}\Lag\right)\right]
\nonumber
\eea
with the same relation as above for the coefficients $g^\eta_{A\alpha}$.
Of course, the normalization condition
$$
\inn({\cal K})[\bar\tau^{1*}\omega\circ{\cal F}\Lag)]=1
$$
 allows to take $F=1$, and selects a representative
on each class $\{{\cal K}\}$ of restricted field operators.
 This implies $\bar\tau^1$-transversality condition for ${\cal K}$.

\begin{quote}
{\bf Remark}:
In the particular case $M\equiv\Real$, the expressions (\ref{coefk}),
(\ref{tildekcor}) and (\ref{kcor}) lead to the local expressions of
the extended and restricted $K$-operators of time-dependent mechanical
systems given in \cite{CFM-95}.
\end{quote}

\section{Properties of the field operators}
\protect\label{pmfo}

\subsection{The Lagrangian equations}

Next we study the properties of the field operators
in relation to the Lagrangian equations.
As we have stated three different but equivalent approaches
to the concept of field operator (namely jet fields,
Ehresmann connection forms or classes of $m$-vector fields
along the Legendre maps), we use the most suitable in each case,
in order to make the proofs easier.

The result (in mechanics) that we want to generalize is relation
(\ref{int000}) (on a submanifold $S\hookrightarrow\Tan Q$).
Thus, we have the following relation
between the field operators and the
solutions of the Lagrangian field equations
(Euler-Lagrange jet fields, multivector fields and connections):

\begin{teor}
Let $\ls$ be a Lagrangian system.
\ben
\item
Let $\widetilde{{\cal Y}_{\cal K}}$ be an extended jet field operator
associated with $\ls$. If there exist a jet field
${\mit\Psi}_{\Lag}\colon J^1E\to J^1J^1E$,
and a $\bar\pi^1$-transverse submanifold
$\j_S\colon S\hookrightarrow J^1E$, such that
\beq
j^1\widetilde{{\cal F}\Lag}\circ{\mit\Psi}_\Lag
\feble{S} \widetilde{{\cal Y}_{\cal K}}
\label{elinduced}
\eeq
then ${\mit\Psi}_\Lag$ is an Euler-Lagrange jet field for $\ls$, on $S$.

Conversely, given an Euler-Lagrange jet field
${\mit\Psi}_\Lag$ for $\ls$, then (\ref{elinduced}) defines an
extended jet field operator $\widetilde{{\cal Y}_{\cal K}}$ for $\ls$, on $S$.
\item
Let $\{\tilde{\cal K}\}$ be
a class of extended $m$-vector field operators associated with $\ls$.
If there exist a class of $m$-vector fields
$\{ X_\Lag\}\subset\vf^m(J^1E)$,
and a $\bar\pi^1$-transverse submanifold
$\j_S\colon S\hookrightarrow J^1E$, such that,
for every $\tilde{\cal K}\in\{\tilde{\cal K}\}$,
\beq
\Lambda^m\Tan\widetilde{{\cal F}\Lag}\circ X_\Lag
\feble{S} \tilde{\cal K}
\quad ;\quad
\mbox{\rm for some $X_\Lag\in\{ X_\Lag\}$}
\label{elinducedc}
\eeq
then $\{ X_\Lag\}$ is a class of
Euler-Lagrange $m$-vector fields for $\ls$, on $S$.

Conversely, given a class of Euler-Lagrange $m$-vector fields
$\{ X_\Lag\}$ for $\ls$, then (\ref{elinducedc}) defines a class of
extended $m$-vector field operators $\{\tilde{\cal K}\}$ for $\ls$, on $S$.
\item
Let $\widetilde{\nabla_{\cal K}}$ be an extended connection operator
associated with $\ls$. If there exist an Ehresmann connection form
$\nabla_{\Lag}\colon J^1E\to\bar\pi^{1*}\Tan^*M\otimes_{J^1E}\Tan J^1E$,
and a $\bar\pi^1$-transverse submanifold
$\j_S\colon S\hookrightarrow J^1E$, such that
\beq
(\tilde\varepsilon_{\Tan^*M}\otimes\Tan\widetilde{{\cal F}\Lag})
\circ\nabla_\Lag \feble{S} \widetilde{\nabla_{\cal K}}
\label{elinducedb}
\eeq
(where
$\tilde\varepsilon_{\Tan^*M}\colon\bar\pi^{1*}\Tan^*M\to
(\bar\tau^1\circ\mu)^*\Tan^*M$
is the natural identification),
then $\nabla_\Lag$ is an Euler-Lagrange connection for $\ls$, on $S$.

Conversely, given an Euler-Lagrange connection
$\nabla_\Lag$ for $\ls$, then (\ref{elinducedb}) defines an
extended jet field operator $\widetilde{\nabla_{\cal K}}$ for $\ls$, on $S$.
\een
\label{equivlag}
\end{teor}
\proof
First, as a guideline for the proof, consider the following  diagram
(which, in general, is not commutative unless restricted to the
appropriate submanifols):
\bea
\begin{array}{ccccc}
& \bar\pi^{1*}\Tan^*M\otimes_{J^1E}\Tan J^1E
&
\begin{picture}(135,20)(0,0)
\put(35,9){\mbox{$\tilde\varepsilon_{\Tan^*M}\otimes
 \Tan\widetilde{{\cal F}\Lag}$}}
\put(0,3){\vector(1,0){135}}
\end{picture}
& (\bar\tau^1\circ\mu)^*\Tan^*M\otimes_{{\cal M}\pi}\Tan{\cal M}\pi &
\\
& \begin{picture}(15,35)(0,0)
\put(0,0){\vector(0,1){35}}
\put(3,12){\mbox{$\Upsilon'_{J^1E}$}}
\end{picture}
& &
\begin{picture}(15,35)(0,0)
\put(10,0){\vector(0,1){35}}
\put(14,12){\mbox{$\Upsilon'_{{\cal M}\pi}$}}
\end{picture} &
\\
\begin{picture}(5,20)(0,0)
\put(-25,0){\mbox{$\{\Lambda^m\Tan J^1E\}$}}
\end{picture}
 & \supset D^m\Tan J^1E
&
\begin{picture}(135,20)(0,0)
\put(15,11){\mbox{$\{\Lambda^m\Tan\widetilde{{\cal F}\Lag}\}$}}
\put(-15,3){\vector(1,0){170}}
\end{picture}
&
D^m\Tan{\cal M}\pi \subset &
\begin{picture}(5,20)(0,0)
\put(-35,0){\mbox{$\{\Lambda^m\Tan{\cal M}\pi\}$}}
\end{picture}
\\
& \begin{picture}(15,35)(0,0)
\put(0,0){\vector(0,1){35}}
\put(3,12){\mbox{$\Upsilon_{J^1E}$}}
\end{picture}
& &
\begin{picture}(15,35)(0,0)
\put(10,0){\vector(0,1){35}}
\put(14,12){\mbox{$\Upsilon_{{\cal M}\pi}$}}
\end{picture}  &
\\
& J^1J^1E
&
\begin{picture}(135,20)(0,0)
\put(-10,11){\mbox{$j^1\widetilde{{\cal F}\Lag}$}}
\put(-15,3){\vector(1,0){170}}
\end{picture}
&
J^1{\cal M}\pi  &
\\
& \begin{picture}(15,35)(0,0)
\put(-40,65){\mbox{$\{ X_\Lag\}$}}
\put(-10,0){\vector(0,1){100}}
\put(0,0){\vector(0,1){35}}
\put(3,12){\mbox{${\mit\Psi}_\Lag$}}
\put(40,0){\vector(0,1){160}}
\put(43,70){\mbox{$\nabla_\Lag$}}
\end{picture}
 &
\begin{picture}(135,35)(0,0)
\put(92,132){\mbox{$\widetilde{\nabla_{\cal K}}$}}
\put(-15,5){\vector(1,1){150}}
\put(72,60){\mbox{$\{\tilde{\cal K}\}$}}
\put(-15,0){\vector(2,1){170}}
\put(60,26){\mbox{$j^1(\tau^1\circ\mu)$}}
\put(155,35){\vector(-4,-1){170}}
\put(105,5){\mbox{$\widetilde{{\cal Y}_{\cal K}}$}}
\put(-15,-12){\vector(4,1){170}}
\end{picture}
&
\begin{picture}(15,35)(0,0)
\put(10,35){\vector(0,-1){35}}
\put(14,12){\mbox{$\pi^1_{{\cal M}\pi}$}}
\end{picture}  &
\\
& J^1E
&
\begin{picture}(135,20)(0,0)
\put(63,11){\mbox{$\widetilde{{\cal F}\Lag}$}}
\put(-15,3){\vector(1,0){170}}
\end{picture}
&
{\cal M}\pi  &

\\
& \begin{picture}(15,35)(0,0)
\put(5,35){\vector(0,-1){35}}
\put(8,12){\mbox{$X_\Lag$}}
\put(-20,0){\vector(-1,3){50}}
\put(-70,70){\mbox{$\rho_{J^1E}$}}
\end{picture}
&
\begin{picture}(135,35)(0,0)
\put(95,13){\mbox{$\tilde{\cal X}$}}
\put(-15,35){\vector(4,-1){170}}
\end{picture}
 &
\begin{picture}(15,35)(0,0)
\put(10,0){\vector(0,1){35}}
\put(14,12){\mbox{$\sigma_{{\cal M}\pi}$}}
\put(35,0){\vector(1,3){50}}
\put(65,70){\mbox{$\rho_{{\cal M}\pi}$}}
\end{picture}  &
\\
& \Lambda^m\Tan J^1E
&
\begin{picture}(135,20)(0,0)
\put(50,9){\mbox{$\Lambda^m\Tan\widetilde{{\cal F}\Lag}$}}
\put(-15,3){\vector(1,0){170}}
\end{picture}
&
\Lambda^m\Tan{\cal M}\pi &
\end{array}
\label{next}
\eea
\ben
\item
We must prove that both the semi-holonomy condition,
and the field equation condition hold for $\widetilde{{\cal Y}_{\cal K}}$
if, and only if, they hold for ${\mit\Psi}_\Lag$.
In this proof all the equalities hold on $S$.

On the one hand, and in relation to the semi-holonomy, we have that
$$
j^1(\tau^1\circ\mu)\circ\widetilde{{\cal Y}_{\cal K}}=
j^1(\tau^1\circ\mu)\circ j^1\widetilde{{\cal F}\Lag}\circ{\mit\Psi}_\Lag=
j^1(\tau^1\circ\mu\circ\widetilde{{\cal F}\Lag})\circ{\mit\Psi}_\Lag=
j^1\pi^1\circ{\mit\Psi}_\Lag
$$
which relates the semi-holonomy of $\widetilde{{\cal Y}_{\cal K}}$
and ${\mit\Psi}_{\Lag}$.

On the other hand, for the field equation we obtain
$$
\widetilde{{\cal F}\Lag}^*[\inn(\widetilde{{\cal Y}_{\cal K}})
(\Omega\circ\widetilde{{\cal F}\Lag})] =
\widetilde{{\cal F}\Lag}^*
[\inn(j^1\widetilde{{\cal F}\Lag}\circ{\mit\Psi}_\Lag)
(\Omega\circ\widetilde{{\cal F}\Lag})]=
\inn({\mit\Psi}_\Lag)(\widetilde{{\cal F}\Lag}^*\Omega)=
\inn({\mit\Psi}_\Lag)\Omega_\Lag
$$
hence the field equation condition holds for $\widetilde{{\cal Y}_{\cal K}}$
if, and only if, the Lagrangian field equation holds for ${\mit\Psi}_\Lag$.
\item
Bearing in mind the commutativity of the diagram (\ref{next})
(on the appropriate submanifolds),
from (\ref{elinduced}) we obtain that
\beann
j^1\widetilde{{\cal F}\Lag}\circ{\mit\Psi}_\Lag
\feble{S} \widetilde{{\cal Y}_{\cal K}}
&\quad \Leftrightarrow\quad&
j^1\widetilde{{\cal F}\Lag}\circ\Upsilon_{J^1E}^{-1}\circ\{ X_\Lag\}
\feble{S}\Upsilon_{{\cal M}\pi}^{-1}\circ\{\tilde{\cal K}\}
\\ &\quad \Leftrightarrow\quad&
\Upsilon_{{\cal M}\pi}\circ j^1\widetilde{{\cal F}\Lag}
\circ\Upsilon_{J^1E}^{-1}\circ\{ X_\Lag\}=
\Lambda^m\Tan\widetilde{{\cal F}\Lag}\circ\{ X_\Lag\}
\feble{S}\{\tilde{\cal K}\}
\eeann
which leads to the relation (\ref{elinducedc}),
for every representative $X_\Lag\in\{ X_\Lag\}$.
As $\{\tilde{\cal K}\}$ is the class associated with
$\widetilde{{\cal Y}_{\cal K}}$, it is made of
semi-holonomic $m$-vector fields along $\widetilde{{\cal F}\Lag}$.
In addition, following the same reasoning as in the above item,
it is proved that all of them verify the field equation condition.
\item
Finally, following the same pattern as in the last item,
(\ref{elinduced}) leads to (\ref{elinducedb}), and
as $\widetilde{\nabla_{\cal K}}$ is associated with
$\widetilde{{\cal Y}_{\cal K}}$, it is semi-holonomic.
In addition, following the same reasoning as in item 1,
it is proved that it verifies the field equation condition.
\qed
\een

\begin{quote}
{\bf Comment}:
Observe that relations (\ref{elinduced}), (\ref{elinducedc})
 and (\ref{elinducedb}) arise because the commutativity of diagram
(\ref{next}) demands it.
\end{quote}

As a straighforward consequence of this Theorem, 
and the Remark after Definition (\ref{restk}), we obtain:

\begin{corol}
If relations (\ref{elinduced}), (\ref{elinducedc}) and (\ref{elinducedb})
hold for the extended field operators, then the following ones hold for
their associated restricted field operators:
\beann
j^1{\cal F}\Lag\circ{\mit\Psi}_\Lag &\feble{S}& {\cal Y}_{\cal K}
\\
\Lambda^m\Tan{\cal F}\Lag\circ X_\Lag &\feble{S}& {\cal K}
\quad ;\quad
\mbox{\rm for every ${\cal K}\in\{{\cal K}\}$,
and for some $X_\Lag\in\{ X_\Lag\}$}
\\
(\varepsilon_{\Tan^*M}\otimes\Tan{\cal F}\Lag)\circ\nabla_\Lag
&\feble{S}& \nabla_{\cal K}
\quad ;\quad
\mbox{\rm (with
$\varepsilon_{\Tan^*M}\colon\bar\pi^{1*}\Tan^*M\to\bar\tau^{1*}\Tan^*M$)}
\eeann
\end{corol}

Then, assuming all these relations, we have:

\begin{teor}
\dst \varphi\colon M\stackrel{\varphi_S}{\longrightarrow}S
\stackrel{\j_S}{\hookrightarrow}J^1E\)
is an integral section of $\{{\cal K}\}$ if, and only if,
it is an integral section of $\{ X_\Lag\}$ too.
Moreover, every integral section $\varphi:=\j_S\circ\varphi_S$
is an holonomic section.
That is, the class $\{{\cal K}\}$,
and its associated ${\cal Y}_{\cal K}$ and $\nabla_{\cal K}$
are integrable if, and only if, the class $\{ X_\Lag\}$,
and its associated ${\cal Y}_\Lag$ and $\nabla_\Lag$ are integrable too.

The same result holds for the extended field operators
$\{\tilde{\cal K}\}$, $\widetilde{{\cal Y}_{\cal K}}$
and $\widetilde{\nabla_{\cal K}}$.
\label{inteor}
\end{teor}
\proof
If $\varphi\colon M\stackrel{\varphi_S}{\longrightarrow}S
\stackrel{\j_S}{\hookrightarrow}J^1E$ is an integral section
of ${\cal K}\in\{{\cal K}\}$ (on $S$), we have
$$
\Lambda^m\Tan({\cal F}\Lag\circ\varphi)=
\Lambda^m\Tan{\cal F}\Lag\circ\Lambda^m\Tan\varphi=
f{\cal K}\circ\varphi\circ\sigma_M=
f(\Lambda^m\Tan{\cal F}\Lag\circ X_\Lag)\circ\varphi\circ\sigma_M
$$
where $f\in\Cinfty (J^1E)$ is a non-vanishing function.
However, if we recall that $\ker\,\Tan_{\varphi(x)}{\cal F}\Lag$
are $\pi^1$-vertical vectors, and then also $\bar\pi^1$-vertical,
from the above equality we can conclude that
$$
\Lambda^m\Tan\varphi=gX_\Lag\circ\varphi\circ\sigma_M
$$
$g\in\Cinfty (J^1E)$ being another non-vanishing function.
So $\varphi:=\j_S\circ\varphi_S$ is an integral section of
 $X_\Lag\in\{ X_\Lag\}$ (on $S$).
The converse is proved by reversing this reasoning.

As $\varphi$ are integral sections of semi-holonomic
$m$-vector fields, they are holonomic sections necessarily.

Finally, the result for the extended field operators is a consequence of
Proposition \ref{insecrestk}.
\qed

{\bf Remarks}:
\bit
\item
If $\ls$ is hyper-regular (recall that, in this case,
$\widetilde{{\cal F}\Lag}$ is a diffeomorphism onto its image,
and ${\cal F}\Lag$ is a diffeomorphism),
 then $S=J^1E$, and the correspondence between field operators and the
corresponding Euler-Lagrange solutions of the Lagrangian field equations
is one-to one.

If $\ls$ is almost-regular, then not only one, but a family of
Euler-Lagrange solutions of the Lagrangian field equations
is associated with every field operator.
\item
In addition, if the integrability condition holds only in a submanifold
${\cal I}\hookrightarrow S$, Theorem \ref{inteor} holds only on ${\cal I}$.

Observe also that this Theorem establishes the  property analogous
to the first one given in Section \ref{eok} for the evolution operator $K$
in mechanics.
\eit

\subsection{The Hamiltonian equations}

Now we will study the properties of the field operators
in relation to the Hamiltonian equations
(see the Section \ref{hf} for the necessary background).

The result (in mechanics) that we want to generalize is relation
(\ref{aux000}) (on a submanifold $S\hookrightarrow\Tan Q$).
Thus, we have the following relation
between the field operators and the
solutions of the Lagrangian field equations
(HDW jet fields, multivector fields and connections):

\begin{teor}
Let $\ls$ be an almost-regular Lagrangian system,
and $\hso$ its associated Hamiltonian system.
\ben
\item
Let $\widetilde{{\cal Y}_{\cal K}}$ be an extended jet field operator
associated with $\ls$. If there exist a jet field
${\mit\Psi}_{{\cal H}_o}\colon{\cal P}\to J^1{\cal P}$,
and a $\bar\pi^1$-transverse submanifold
$\j_S\colon S\hookrightarrow J^1E$,
 such that
\beq
j^1\tilde\jmath_0\circ j^1\tilde h\circ{\mit\Psi}_{{\cal H}_o}
\circ{\cal F}\Lag_0 \feble{S} \widetilde{{\cal Y}_{\cal K}}
\label{elinduced2}
\eeq
then ${\mit\Psi}_{{\cal H}_o}$ is a Hamilton-De Donder-Weyl
 jet field for $\hso$, on $P={\cal F}\Lag (S)$.

Conversely, given a Hamilton-De Donder-Weyl jet field
${\mit\Psi}_{{\cal H}_o}$ for $\hso$, on a submanifold
$P\hookrightarrow{\cal P}$, then (\ref{elinduced2}) defines a
jet field $\widetilde{{\cal Y}_{\cal K}}$ along $\widetilde{{\cal F}\Lag}$,
on every submanifold $S\hookrightarrow J^1E$
such that ${\cal F}\Lag(S)=P$,
which satisfy the structural and the field equation conditions of
Definition \ref{tildek}, but not the semi-holonomy condition necessarily.
\item
Let $\{\tilde{\cal K}\}$ be
a class of extended field operators associated with $\ls$.
If there exist a class of $m$-vector fields
$\{ X_{{\cal H}_o}\}\subset\vf^m({\cal P})$,
and a $\bar\pi^1$-transverse submanifold
$\j_S\colon S\hookrightarrow J^1E$, such that,
for every $\tilde{\cal K}\in\{\tilde{\cal K}\}$,
\beq
\Lambda^m\Tan\tilde\jmath_0\circ\Lambda^m\Tan\tilde h\circ
X_{{\cal H}_o}\circ{\cal F}\Lag_0 \feble{S}\tilde{\cal K}
\quad ;\quad
\mbox{\rm for some $X_{{\cal H}_o}\in\{ X_{{\cal H}_o}\}$}\, ,
\label{elinduced2c}
\eeq
then $\{ X_{{\cal H}_o}\}$ is a class of
Hamilton-De Donder-Weyl $m$-vector fields for $\hso$, on $P={\cal F}\Lag (S)$.

Conversely, if $\{ X_{{\cal H}_o}\}$ is a class of
Hamilton-De Donder-Weyl $m$-vector fields for $\hso$, on a submanifold
$P\hookrightarrow{\cal P}$, then 
the above relation defines a class of
 $m$-vector fields $\{\tilde{\cal K}\}$ along $\widetilde{{\cal F}\Lag}$,
on every submanifold $S\hookrightarrow J^1E$
such that ${\cal F}\Lag(S)=P$,
which satisfy the structural and the field equation conditions of
 Definition \ref{tildek}, but not the semi-holonomy condition necessarily.
\item
Let $\widetilde{\nabla_{\cal K}}$ be an extended connection operator
associated with $\ls$. If there exist an Ehresmann connection form
$\nabla_{{\cal H}_o}\colon{\cal P}\to
\bar\tau_0^{1*}\Tan^*M\otimes_{\cal P}\Tan{\cal P}$,
and a $\bar\pi^1$-transverse submanifold
$\j_S\colon S\hookrightarrow J^1E$, such that
\beq
(\tilde\varepsilon^0_{\Tan^*M}\otimes\Tan\widetilde{{\cal F}\Lag}_0)
\circ\nabla_{{\cal H}_o}\circ{\cal F}\Lag_0
 \feble{S} \widetilde{\nabla_{\cal K}}
\label{elinduced2b}
\eeq
(where
$\tilde\varepsilon^0_{\Tan^*M}\colon\bar\tau_0^{1*}\Tan^*M\to
(\bar\tau^1\circ\mu)^*\Tan^*M$
is the natural identification),
then $\nabla_{{\cal H}_o}$ is a Hamilton-De Donder-Weyl connection for $\hso$,
on $P={\cal F}\Lag (S)$.

Conversely, given a Hamilton-De Donder-Weyl connection
$\nabla_{{\cal H}_o}$ for $\hso$, on a submanifold
$P\hookrightarrow{\cal P}$, then the above relation defines an
Ehresmann connection form $\widetilde{\nabla_{\cal K}}$
 along $\widetilde{{\cal F}\Lag}$,
on every submanifold $S\hookrightarrow J^1E$
such that ${\cal F}\Lag(S)=P$,
which satisfy the structural and the field equation conditions of
 Definition \ref{tildek}, but not the semi-holonomy condition necessarily.
\een
If $\ls$ is a hyper-regular Lagrangian system,
and $\hs$ its associated Hamiltonian system,
then the same results hold (with $S=J^1E$).
In the converse statements however,
the jet field $\widetilde{{\cal Y}_{\cal K}}$, Ehresmann connection form
$\widetilde{\nabla_{\cal K}}$, and classes of
$m$-vector fields $\{\tilde{\cal K}\}$ along $\widetilde{{\cal F}\Lag}$
also satisfy also the semiholonomy condition, and hence they
are extended field operators for $\ls$.
\label{equivham}
\end{teor}
\proof
First we prove item 2.
As a standpoint, consider the following diagram,
(which, in general, is not commutative unless restricted to the
appropriate submanifols)
(see also diagram (\ref{arhs})):
\bea
\begin{array}{ccccccc}
\Lambda^m\Tan{\cal M}\pi
 &
\begin{picture}(40,20)(0,0)
\put(12,8){\mbox{$\sigma_{{\cal M}\pi}$}}
\put(0,3){\vector(1,0){40}}
\end{picture}
 &
 {\cal M}\pi &
\begin{picture}(40,20)(0,0)
\put(60,8){\mbox{$\mu$}}
\put(0,3){\vector(1,0){125}}
\end{picture}
& & & J^{1*}E
\\
& & &
\begin{picture}(40,40)(0,0)
\put(28,20){\mbox{$\widetilde{{\cal F}\Lag}$}}
\put(40,0){\vector(-1,1){40}}
\put(-64,13){\mbox{$\tilde{\cal K}$}}
\put(40,-6){\vector(-3,1){130}}
\end{picture}
& &
\begin{picture}(40,40)(0,0)
\put(-5,20){\mbox{${\cal F}\Lag$}}
\put(0,0){\vector(1,1){40}}
\end{picture}
&
\\
& & & & J^1E & &
\\
& &
\begin{picture}(10,40)(0,0)
\put(-10,45){\mbox{$\tilde\jmath_0$}}
\put(5,0){\vector(0,1){95}}
\end{picture}
 &
\begin{picture}(40,40)(0,0)
\put(-5,20){\mbox{$\widetilde{{\cal F}\Lag_0}$}}
\put(40,40){\vector(-1,-1){40}}
\end{picture}
& &
\begin{picture}(40,40)(0,0)
\put(25,20){\mbox{${\cal F}\Lag_0$}}
\put(0,40){\vector(1,-1){40}}
\end{picture}
&
\begin{picture}(10,40)(0,0)
\put(8,45){\mbox{$\jmath_0$}}
\put(5,0){\vector(0,1){95}}
\end{picture}
\\
& &
 \tilde{\cal P}
 &
\begin{picture}(40,10)(0,0)
\put(60,13){\mbox{$\tilde\mu$}}
\put(0,8){\vector(1,0){125}}
\put(47,-10){\mbox{$\tilde\mu^{-1}\equiv\tilde h$}}
\put(125,3){\vector(-1,0){125}}
\end{picture}
& & & {\cal P}
\\
\begin{picture}(10,40)(0,0)
\put(-30,75){\mbox{$\Lambda^m\Tan\tilde\jmath_0$}}
\put(5,0){\vector(0,1){150}}
\end{picture}
 &
\begin{picture}(40,40)(0,0)
\put(23,15){\mbox{$\sigma_{\tilde{\cal P}}$}}
\put(0,0){\vector(1,1){40}}
\end{picture}
 & & & & &
\begin{picture}(10,40)(0,0)
\put(-22,15){\mbox{$X_{{\cal H}_o}$}}
\put(3,40){\vector(0,-1){40}}
\put(11,15){\mbox{$\sigma_{{\cal P}}$}}
\put(8,0){\vector(0,1){40}}
\end{picture}
\\
\Lambda^m\Tan\tilde{\cal P}
&
\begin{picture}(40,10)(0,0)
\put(100,7){\mbox{$\Lambda^m\Tan\tilde h$}}
\put(200,3){\vector(-1,0){200}}
\end{picture}
& & & & &
\Lambda^m\Tan{\cal P}
\end{array}
\label{next2}
\eea
Then we have that
\beann
\widetilde{{\cal F}\Lag}^*
[\inn(\tilde{\cal K})(\Omega\circ\widetilde{{\cal F}\Lag})]&=&
(\tilde\jmath_0\circ\tilde h\circ{\cal F}\Lag_0)^*[
\inn(\Lambda^m\Tan\tilde\jmath_0\circ\Lambda^m\Tan\tilde h\circ X_{{\cal H}_o}
\circ{\cal F}\Lag_0)(\Omega\circ\widetilde{{\cal F}\Lag})]
\\ &=&
{\cal F}\Lag_0^*\{(\tilde\jmath_0\circ \tilde h)^*
[\inn(\Lambda^m\Tan(\tilde\jmath_0\circ\tilde h)\circ
 X_{{\cal H}_o}\circ{\cal F}\Lag_0)(\Omega\circ\widetilde{{\cal F}\Lag})]\}=
{\cal F}\Lag_0^*[\inn(X_{{\cal H}_o})\Omega_h^0]
\eeann
where all the equalities hold on $S$.
But, as ${\cal F}\Lag_0$ is a submersion, we obtain that
$$
\widetilde{{\cal F}\Lag}^*
[\inn(\tilde{\cal K})(\Omega\circ\widetilde{{\cal F}\Lag})]
\feble{S}0
\ \Longleftrightarrow \
\inn(X_{{\cal H}_o})\Omega_h^0 \feble{P} 0
$$
hence the field equation condition holds for $\tilde{\cal K}$
on $S$ if, and only if, the Hamiltonian field equation
 holds for $X_{{\cal H}_o}$ on $P$.

The proof of items 1 and 3 follow the same pattern as
the proof of items 2 and 3 of Theorem \ref{equivlag}.

For hyper-regular systems, the proof of these properties is the same,
but taking into acount that now ${\cal P}=J^{1*}E$,
${\cal F}\Lag_0={\cal F}\Lag$, and $h=\tilde\jmath_0\circ\tilde h$.
In addition,  the classes of
HDW $m$-vector fields, HDW jet fields, and HDW connections
are defined everywhere in $J^{1*}E$. Thus,
the only addendum is to prove that, if
$X_{\cal H}$ is a Hamilton-De Donder-Weyl
$m$-vector field for $\hs$, then its associated
 $m$-vector field along $\widetilde{{\cal F}\Lag}$, $\tilde{\cal K}$,
is semi-holonomic. As $X_{\cal H}\in\vf^M(J^{1*}E)$, by definition
$\sigma_{J^{1*}E}\circ X_{\cal H}={\rm Id}_{J^{1*}E}$,
then, recalling the definition of the map
$\tilde\varrho_E\colon\Lambda^m\Tan{\cal M}\pi\to J^1E$
(see (\ref{tilderho})),
and taking into account that ${\cal F}\Lag$ is a diffeomorphism, we have that
$$
\tilde\varrho_E\circ\tilde{\cal K}=
\tilde\varrho_E\circ\Lambda^m\Tan h\circ X_{\cal H}\circ{\cal F}\Lag=
{\cal F}\Lag^{-1}\circ\sigma_{J^{1*}E}\circ X_{\cal H}\circ{\cal F}\Lag=
{\rm Id}_{J^1E}
$$
which is the condition for $\tilde{\cal K}$
(and hence for $\widetilde{{\cal Y}_{\cal K}}$
 and $\widetilde{\nabla_{\cal K}}$) to be semi-holonomic.
That is, we have the following diagram
$$
\begin{array}{ccccc}
\Lambda^m\Tan J^{1*}E &
\begin{picture}(40,20)(0,0)
\put(52,8){\mbox{$\Lambda^m\Tan h$}}
\put(0,3){\vector(1,0){125}}
\end{picture}
& & & \Lambda^m\Tan{\cal M}\pi
\\
 & & &
\begin{picture}(40,40)(0,0)
\put(20,5){\mbox{$\tilde{\cal K}$}}
\put(33,40){\vector(-1,-1){40}}
\put(-10,25){\mbox{$\tilde\varrho_E$}}
\put(0,0){\vector(1,1){40}}
\end{picture}
&
\\
 & & J^1E & &
\\
\begin{picture}(10,40)(0,0)
\put(10,45){\mbox{$X_{\cal H}$}}
\put(7,0){\vector(0,1){95}}
\put(-30,45){\mbox{$\sigma_{J^{1*}E}$}}
\put(3,95){\vector(0,-1){95}}
\end{picture}
 &
\begin{picture}(40,40)(0,0)
\put(0,20){\mbox{${\cal F}\Lag$}}
\put(40,40){\vector(-1,-1){40}}
\end{picture}
& &
\begin{picture}(40,40)(0,0)
\put(25,20){\mbox{$\widetilde{{\cal F}\Lag}$}}
\put(0,40){\vector(1,-1){40}}
\end{picture}
&
\begin{picture}(10,40)(0,0)
\put(8,45){\mbox{$\sigma_{{\cal M}\pi}$}}
\put(5,0){\vector(0,1){95}}
\end{picture}
\\
 J^{1*}E
 &
\begin{picture}(40,10)(0,0)
\put(60,10){\mbox{$h$}}
\put(0,3){\vector(1,0){125}}
\end{picture}
& & & {\cal M}\pi
\end{array}
$$
\qed

\begin{quote}
{\bf Comment}:
Observe that relations (\ref{elinduced2}), (\ref{elinduced2c})
 and (\ref{elinduced2b}) arise because the commutativity of diagram
(\ref{next2}) demands it.
\end{quote}

As a straighforward consequence of this Theorem, and the Remark after
 Definition (\ref{restk}), we obtain:

\begin{corol}
If the relations (\ref{elinduced2}), (\ref{elinduced2c})
 and (\ref{elinduced2b}) hold for the extended field operators,
then the following ones hold for their associated
restricted field operators:
\beann
j^1\jmath_0\circ{\mit\Psi}_{{\cal H}_o}
\circ{\cal F}\Lag_0 &\feble{S}& {\cal Y}_{\cal K}
\\
\Lambda^m\Tan\jmath_0\circ
X_{{\cal H}_o}\circ{\cal F}\Lag_0 &\feble{S}& {\cal K}
\quad ;\quad
\mbox{\rm  for every ${\cal K}\in\{{\cal K}\}$,
and for some $X_{{\cal H}_o}\in\{ X_{{\cal H}_o}\}$}
 \\
(\varepsilon^0_{\Tan^*M}\otimes\Tan{\cal F}\Lag_0)
\circ\nabla_{{\cal H}_o}\circ{\cal F}\Lag_0
 &\feble{S}& \nabla_{\cal K}
\quad , \quad
\mbox{\rm (with
$\varepsilon^0_{\Tan^*M}\colon\bar\tau^{1*}_0\Tan^*M\to\bar\tau^{1*}\Tan^*M$)}
\eeann
(with the same restrictions in relation to the semi-holonomy condition).
\end{corol}

Then assuming all these relations, we have:

\begin{teor}
The class $\{{\cal K}\}$, and its associated
${\cal Y}_{\cal K}$ and $\nabla_{\cal K}$
are integrable if, and only if, the class 
$\{ X_{{\cal H}_o}\}$, and its associated ${\cal Y}_{{\cal H}_o}$
 and $\nabla_{{\cal H}_o}$ are integrable too. In particular:
\ben
\item
Let ${\cal F}\Lag_S\colon S\to P$ be the restriction of
${\cal F}\Lag_0$ to $S$ (that is, 
$\jmath_0\circ{\cal F}\Lag_S={\cal F}\Lag_0\circ\j_S$).
If \dst\varphi\colon M\stackrel{\varphi_S}{\longrightarrow}S
\stackrel{\j_S}{\hookrightarrow}J^1E\)
 is an integral section of $\{{\cal K}\}$ on $S$, then
\dst\psi_o\colon M\stackrel{\psi_P}{\longrightarrow}P
\stackrel{\jmath_P}{\hookrightarrow}{\cal P}\) 
 is an integral section of $\{ X_{{\cal H}_o}\}$ on $P$,
where $\psi_P:={\cal F}\Lag_S\circ\varphi_S$.
\item
 Conversely, if 
\dst\psi_o\colon M\stackrel{\psi_P}{\longrightarrow}P
\stackrel{\jmath_P}{\hookrightarrow}{\cal P}\)
is an integral section of $\{ X_{{\cal H}_o}\}$ on $P$, then the section
\dst\varphi\colon M\stackrel{\varphi_S}{\longrightarrow}S
\stackrel{\j_S}{\hookrightarrow}J^1E\) ,
is an integral section of $\{{\cal K}\}$ on $S$,
for every $\varphi_S\colon M\to S\subseteq J^1E$
such that $\psi_P={\cal F}\Lag_S\circ\varphi_S$.

The section $\varphi_S$, and hence $\varphi:=\j_S\circ\varphi_S$,
are holonomic if, and only if, the class $\{{\cal K}\}$
is semi-holonomic (and hence is a class of field operators).
\een

The same result holds for the extended field operators
$\{\tilde{\cal K}\}$, $\widetilde{{\cal Y}_{\cal K}}$
and $\widetilde{\nabla_{\cal K}}$.
\label{inteorbis}
\end{teor}
\proof
If the system is almost-regular, consider the diagram
\beq
\begin{array}{ccccc}
&
\begin{picture}(80,25)(0,0)
\put(75,10){\mbox{$\Lambda^m\Tan({\cal F}\Lag\circ\varphi)$}}
\put(0,5){\vector(1,0){210}}
\end{picture}
& & &
\\
\Lambda^m\Tan M &
\begin{picture}(80,10)(0,0)
\put(3,7){\mbox{$\Lambda^m\Tan({\cal F}\Lag_0\circ\varphi)$}}
\put(0,3){\vector(1,0){80}}
\end{picture}
 & \Lambda^m\Tan{\cal P} &
\begin{picture}(80,10)(0,0)
\put(20,6){\mbox{$\Lambda^m\Tan\jmath_0$}}
\put(0,3){\vector(1,0){80}}
\end{picture} &
\Lambda^m\Tan J^{1*}E
\\
\begin{picture}(10,60)(0,0)
\put(8,25){\mbox{$\sigma_M$}}
\put(5,60){\vector(0,-1){60}}
\end{picture}
 & &
\begin{picture}(10,60)(0,0)
\put(-20,40){\mbox{$X_{{\cal H}_o}$}}
\put(5,0){\vector(0,1){60}}
\end{picture}
& &
\begin{picture}(10,60)(0,0)
\put(8,25){\mbox{$\sigma_{J^{1*}E}$}}
\put(5,60){\vector(0,-1){60}}
\end{picture}
\\
 M &
\begin{picture}(80,10)(0,0)
\put(3,-4){\mbox{$\varphi$}}
\put(-10,6){\vector(1,0){37}}
\put(32,0){\mbox{$J^1E$}}
\put(70,-5){\mbox{${\cal F}\Lag_0$}}
\put(65,6){\vector(1,0){27}}
\put(150,40){\mbox{${\cal K}$}}
\put(53,19){\vector(3,1){160}}
\put(15,20){\mbox{$\psi_0$}}
\put(-10,13){\vector(1,0){100}}
\put(-10,-9){\vector(2,-1){105}}
\end{picture}
 & {\cal P} &
\begin{picture}(80,10)(0,0)
\put(35,9){\mbox{$\jmath_0$}}
\put(0,3){\vector(1,0){80}}
\end{picture}
 &
J^{1*}E
\\
&
\begin{picture}(80,22)(0,0)
\put(135,2){\mbox{${\cal F}\Lag$}}
\put(53,15){\vector(1,0){160}}
\end{picture}
 & & &
\\
  &
\begin{picture}(80,50)(0,0)
\put(-10,30){\mbox{$\varphi_S$}}
\put(-24,67){\vector(1,-1){50}}
\put(37,0){\mbox{$S$}}
\put(27,33){\mbox{$\j_S$}}
\put(65,40){\mbox{$\psi_P$}}
\put(40,15){\vector(0,1){58}}
\put(65,7){\mbox{${\cal F}\Lag_S$}}
\put(55,3){\vector(1,0){40}}
\end{picture}
 &
\begin{picture}(10,50)(0,0)
\put(3,0){\mbox{$P$}}
\put(10,35){\mbox{$\jmath_P$}}
\put(6,15){\vector(0,1){45}}
\end{picture}
 & &
\end{array}
\label{diagfin}
\eeq
(where $X_{{\cal H}_o}$ denotes any extension of the
HDW $m$-vector field solution on $P$ to ${\cal P}$).
\ben
\item
 From (\ref{int1}), $\varphi_S\colon M\to J^1E$ is an integral section
of ${\cal K}\in\{{\cal K}\}$ on $S$ if, and only if,
\beq
\Lambda^m\Tan (\jmath_0\circ{\cal F}\Lag_0\circ\varphi)=
\Lambda^m\Tan ({\cal F}\Lag\circ\j_S\circ\varphi_S)=
f{\cal K}\circ\j_S\circ\varphi_S\circ\sigma_M
\label{integ}
\eeq
Then on the one hand we have that
\beann
\Lambda^m\Tan({\cal F}\Lag\circ\j_S\circ\varphi_S)&=&
\Lambda^m\Tan\jmath_0\circ
\Lambda^m\Tan({\cal F}\Lag_0\circ\j_S\circ\varphi_S)
\\ &=&
\Lambda^m\Tan\jmath_0\circ
\Lambda^m\Tan(\jmath_P\circ{\cal F}\Lag_S\circ\varphi_S)=
\Lambda^m\Tan\jmath_0\circ
\Lambda^m\Tan(\jmath_P\circ\psi_P)
\eeann
and on the other hand,
\beann
f{\cal K}\circ\j_S\circ\varphi_S\circ\sigma_M&=&
f\Lambda^m\Tan\jmath_0\circ
X_{{\cal H}_o}\circ{\cal F}\Lag_0\circ\j_S\circ\varphi_S\circ\sigma_M
\\ &=&
f\Lambda^m\Tan\jmath_0\circ
X_{{\cal H}_o}\circ\jmath_P\circ{\cal F}\Lag_S\circ\varphi_S\circ\sigma_M=
f\Lambda^m\Tan\jmath_0\circ
X_{{\cal H}_o}\circ\jmath_P\circ\psi_P\circ\sigma_M
\eeann
where $f\in\Cinfty (J^1E)$ is a non-vanishing function.
Hence, as $\jmath_0$ is an imbedding,
we obtain that (\ref{integ}) is equivalent to
\beq
\Lambda^m\Tan\psi_0=
\Lambda^m\Tan(\jmath_P\circ\psi_P)=
fX_{{\cal H}_o}\circ\jmath_P\circ\psi_P\circ\sigma_M
\label{integb}
\eeq
which is the condition for $\psi_P$ to be an integral section
of $X_{{\cal H}_o}$ on $P$.
\item
The converse is proved by reversing the above reasoning.
In addition, the sections $\varphi_S$ and $\varphi:=\j_S\circ\varphi_S$
are holonomic if, and only if, they are integral sections
of semi-holonomic $m$-vector fields along the Legendre map.
\een
If the system is hyper-regular the proof is analogous, but taking
${\cal P}=J^{1*}E$ and ${\cal F}\Lag_0={\cal F}\Lag$.

Finally, the result for the extended field operators is a consequence of
Proposition \ref{insecrestk}.
\qed

And as an immediate corollary of this Theorem, we obtain the following
characterization for the Hamiltonian sections:

\begin{teor}
$\psi_0$ is a section solution of the
Hamiltonian problem if, and only if,
the following relation holds for
$\psi:=\jmath_0\circ\psi_0=\jmath_0\circ\jmath_P\circ\psi_P$:
$$
\Lambda^m\Tan\psi=
f{\cal K}\circ j^1(\tau^1\circ\psi)\circ\sigma_M
$$
or, what is equivalent, if
${\cal K}=\Lambda^m\Tan\mu\circ\tilde{\cal K}$,
for $\tilde\psi:=\tilde\jmath_0\circ\tilde h\circ\psi_0$ we have
$$
\Lambda^m\Tan\tilde\psi=
f\tilde{\cal K}\circ j^1(\tau^1\circ\mu\circ\tilde\psi)\circ\sigma_M
$$
\label{inteorbibis}
\end{teor}
\proof
Bearing in mind the commutativity of diagram (\ref{diagfin}),
and taking into account that (\ref{integb})
is the n.s.c. for $\psi_0$ to be an integral section
of $X_{{\cal H}_o}$, we have that
\beann
\Lambda^m\Tan\psi&=&
\Lambda^m\Tan(\jmath_0\circ\psi_0)=
f\Lambda^m\Tan\jmath_0 X_{{\cal H}_o}\circ\psi_0\circ\sigma_M=
f\Lambda^m\Tan\jmath_0\circ X_{{\cal H}_o}
\circ{\cal F}\Lag_0\circ\varphi\circ\sigma_M
\\ &=&
f{\cal K}\circ\varphi\circ\sigma_M=
f{\cal K}\circ j^1\phi\circ\sigma_M=
f{\cal K}\circ j^1(\tau^1\circ\psi)\circ\sigma_M
\eeann
since $\varphi$ is holonomic and, by construction,
$\phi=\tau^1_0\circ\psi_0=\tau^1\circ\psi$, as the following diagram shows
$$
\begin{array}{ccc}
J^1E &
\begin{picture}(135,20)(0,0)
\put(15,8){\mbox{${\cal F}\Lag_0$}}
\put(0,3){\vector(1,0){58}}
\put(63,0){\mbox{${\cal P}$}}
\put(96,8){\mbox{$\jmath_0$}}
\put(80,3){\vector(1,0){58}}
\end{picture}
& J^{1*}E
\\ &
\begin{picture}(135,100)(0,0)
\put(32,84){\mbox{$\pi^1$}}
\put(93,82){\mbox{$\tau^1$}}
\put(0,55){\mbox{$j^1\phi$}}
 \put(118,55){\mbox{$\psi$}}
\put(75,40){\mbox{$\psi_0$}}
 \put(58,30){\mbox{$\phi$}}
 \put(52,80){\mbox{$\tau_0^1$}}
\put(60,55){\mbox{$E$}}
 \put(65,0){\mbox{$M$}}
\put(67,100){\vector(0,-1){35}}
\put(0,100){\vector(3,-2){55}}
 \put(135,100){\vector(-3,-2){55}}
\put(55,13){\vector(-2,3){55}}
 \put(80,13){\vector(2,3){55}}
\put(67,13){\vector(0,1){35}}
 \put(73,13){\vector(0,1){88}}
\end{picture} &
\end{array}
$$
The relation involving $\tilde{\cal K}$ is immediate.
\qed

{\bf Remarks}:
\bit
\item
In both the almost-regular and hyper-regular cases,
 the correspondence between HDW solutions of the Hamiltonian equations
and the corresponding type of extended field operator is one-to-one.
\item
In addition, if the integrability condition holds only in a submanifold
${\cal I}\hookrightarrow S$, then
 Theorems \ref{inteorbis} and \ref{inteorbibis} only holds on ${\cal I}$ and
${\cal F}\Lag ({\cal I})$
(which is assumed to be a submanifold of $P$).
\item
Observe also that Theorem \ref{inteorbis}, together with Theorem
\ref{inteor}, establishes the equivalence between
the Lagrangian and Hamiltonian formalisms.

In its turn, Theorem \ref{inteorbibis} establishes the analogous property
to (\ref{aux00}) for the evolution operator $K$ in mechanics.
\eit

In the light of these results,
the existence of a multiplicity of extended field operators
is hardly surprising, since in Lagrangian and
Hamiltonian field theories, solutions of the
field equations are not unique.

 \section{Conclusions and outlook}

The generalization of the so-called {\sl evolution operator}
of the autonomous mechanical systems to field theories is achieved.
Our geometric framework is the multisymplectic jet bundle description
of these theories.

First, the geometric characteristics of multivector fields, jet fields and
connection forms along maps are stated
(in the jet bundle context which is of interest to us)
as a generalization of the corresponding ones for
multivector fields, jet fields and
connection forms in jet bundles. In particular, the existence of
one-to-one correspondences between  {\it (a)\/}the sets of
equivalence classes of non-vanishing, locally decomposable and transverse 
$m$-vector fields along the (extended) Legendre map,
and {\it (b)\/}the sets of orientable jet fields and orientable connection forms
along this map, is proved and used for establishing
some geometrical characteristics of these objects.

 In this way, the {\sl extended field operators}
are defined as {\sl sections along the extended Legendre map}
in three equivalent ways: as non-vanishing, locally-decomposable 
and transverse  $m$-vector fields;
as orientable jet fields; and as orientable connection forms along this map;
all of them satisfying the conditions of being both semi-holonomic and
a solution of a suitable field equation, which involves the
multisymplectic canonical structure of the extended multimomentum bundle
of the theory.
The existence of these field operators
is proved and, as a relevant difference to mechanics,
we also see that they are not uniquely determined
(and these results do not depend on the regularity of the system).
This is not a surprising fact, since solutions of field equations
are not unique, even for regular field theories.

Furthermore, the so-called {\sl restricted field operators}
are also defined starting from the extended ones. So
they are non-vanishing, locally-decomposable 
and transverse  $m$-vector fields, orientable jet fields, 
or orientable connection forms along the restricted Legendre map
${\cal F}\Lag$.

As the first properties of the field operators, we show
how solutions of the Euler-Lagrange and Hamiltonian field equations
(jet fields, multivector fields and connections)
can be generated from these field operators; and conversely,
starting from these solutions the field operators can be recovered.
In particular, we prove that the integral sections of the field operators
are the section solutions of the Euler-Lagrange equations,
whereas their images by the Legendre map are the integral sections of the 
Hamilton-De Donder-Weyl equations.
Furthermore, it is showed that these
integral sections of the Hamilton-De Donder-Weyl equations
can be characterized using only the field operators. 
Of course, all these relations hold on the submanifolds
where solutions of field equations exist.
All these results establish the relationship between the solutions of field
equations in the Lagrangian and
Hamiltonian formalisms for (singular) field theories.

In this way, the field operators are very
efficient tools to unify the Lagrangian and Hamiltonian
formalisms.

It is interesting to point out that our 
field operators are covariant objects,
and hence they are not ``evolution operators'' in any sense.
In order to define these evolution operators a previous {\sl space-time}
decomposition must be carried out on the base manifold $M$ of the theory.

In further research works these field operators
will be used for carrying out a deeper analysis
of properties of these theories, in the same way as the
evolution operator is used in mechanics.
For instance, either to set the complete relation between the
Lagrangian and Hamiltonian constraint algorithms arising
for almost-regular field theories, or to study the existence
and characterization of symmetries.

\subsection*{Acknowledgments}

We acknowledge the financial support of the CICYT BFM2002-03493.
We wish to thank Mr. Jeff Palmer for his
assistance in preparing the English version of the manuscript.
We are also very grateful to Prof. X. Gr\`acia
for his enlightening comments and suggestions.
Thanks to the referees for their constructive comments.

\end{document}